\documentclass{aa}

\usepackage[T1]{fontenc}
\usepackage[latin1]{inputenc}
\usepackage{wasysym}
\usepackage{graphicx}
\usepackage{natbib}
\usepackage{txfonts}

\newcommand{\eq}[1]{\begin{equation}  #1 \end{equation}}

\newcommand{\eqa}[1]{\begin{eqnarray}   #1 \end{eqnarray}}

\newcommand{\br}[1]{\left( #1 \right)}
\newcommand{\bc}[1]{\left\{ #1 \right\}}
\newcommand{\bb}[1]{\left[ #1 \right]}
\newcommand{\ba}[1]{\left\langle #1 \right\rangle}

\newcommand{\nn}{\nonumber}

\newcommand{\dd}{{\rm d}}
\newcommand{\expo}[1]{~{\rm e}^{ #1 }}
\newcommand{\vek}[1]{\mbox{\boldmath $#1$}}
\newcommand{\ic}{{\rm i}}

\begin{document}

\title{Simultaneous 
measurement 
of cosmology and intrinsic alignments using
joint cosmic shear and galaxy number density correlations}
\titlerunning{Simultaneous 
measurement of cosmic shear and galaxy number density correlations 
}

\author{B. Joachimi\inst{1,2} \and S. L. Bridle\inst{1}} 

\institute{
Department of Physics and Astronomy, University College London, London WC1E 6BT, UK
\and
Argelander-Institut f\"ur Astronomie (AIfA), Universit\"at Bonn, Auf dem H\"ugel 71, 53121 Bonn, Germany\\
\email{joachimi@astro.uni-bonn.de}
}

\date{Received 12 November 2009 / Accepted 9 August 2010}

\abstract{}
{Cosmic shear is a powerful method to constrain cosmology, provided that any systematic effects are under control. The intrinsic alignment of galaxies is expected to severely bias parameter estimates if not taken into account. We explore the potential of a joint analysis of tomographic galaxy ellipticity, galaxy number density, and ellipticity-number density cross-correlations to simultaneously constrain cosmology and
self-calibrate 
unknown intrinsic alignment and galaxy bias contributions.}
{We treat intrinsic alignments and galaxy biasing as free functions of scale and redshift and marginalise over the
resulting parameter sets. 
Constraints on cosmology are calculated by combining
the likelihoods from 
all two-point correlations between galaxy ellipticity and galaxy number density.
The information required for these calculations is already available in a standard cosmic shear data set. 
We include contributions to these functions from cosmic shear, intrinsic alignments, galaxy clustering and magnification effects.}
{In a Fisher matrix analysis we compare our constraints with those from cosmic shear alone in the absence of intrinsic alignments. For a
potential future large area survey, such as Euclid, 
the extra information from the additional correlation functions can make up for the additional free parameters in the intrinsic alignment and galaxy bias terms, depending on the flexibility in the models. For example, the Dark Energy Task Force figure of merit is recovered even when more than 100 free parameters are marginalised over. We find that the redshift quality requirements are similar to those calculated in the absence of intrinsic alignments.}
{}


\maketitle

\section{Introduction}
\label{sec:intro}

On its way to Earth the light
from 
distant galaxies is continuously deflected by the matter density inhomogeneities which it passes by.
This induces distortions of the shapes of the projected galaxy images on the sky, causing modifications of these shapes of order $1\,\%$,
known as weak gravitational lensing, or cosmic shear. 
Hence, to measure this distortion effect on distant galaxies, one requires statistical methods;
see e.g. \citet{bartelmann01} and \citet{schneider06} for detailed reviews.

Detected in 2000 \citep{bacon00,kaiser00,vwaer00,wittman00}, cosmic shear has since rapidly evolved into a mature technique that produces increasingly stringent constraints on cosmological parameters \citep[see e.g.][]{jarvis06,hoekstra06,semboloni06,hetterscheidt07,benjamin07,massey07,schrabback07,fu07,schrabback09}. Probing both the evolution of structure and the geometry of the Universe at low redshifts, it is complementary to observations of the cosmic microwave background
\citep[e.g.][]{tereno05,das08} 
and considered as 
potentially 
the most powerful method to pin down the properties of dark energy \citep{hu02b,albrecht06,peacock06}. Upcoming and future surveys like Pan-STARRS\footnote{Panoramic Survey Telescope \& Rapid Response System, \texttt{http://pan-starrs.ifa.hawaii.edu}}, DES\footnote{Dark Energy Survey, \texttt{https://www.darkenergysurvey.org}}, LSST\footnote{Large Synoptic Survey Telescope, \texttt{http://www.lsst.org}}, JDEM\footnote{Joint Dark Energy Mission, \texttt{http://jdem.gdfc.nasa.gov}}, and Euclid\footnote{\texttt{http://sci.esa.int/science-e/www/area/\\index.cfm?fareaid=102}} will produce deep imaging over large fractions of the sky and thereby yield unprecedented insight into the properties of dark matter, dark energy and gravitation
\citep[e.g.][]{takada04,refregier04,refregier06,kitching08b,thomas08}. 

The
great 
statistical power of cosmic shear demands a 
careful 
assessment of possible systematic errors that might bias the results. A serious limitation may arise from a physical systematic caused by the intrinsic alignment of galaxies. The matter structure around galaxies can modify their intrinsic shape and their orientation. Firstly, this can result in correlations between the intrinsic shapes of galaxies which are close both on the sky and in redshift (intrinsic ellipticity correlations, or II correlations). Moreover, a dark matter halo can intrinsically align a physically close galaxy in the foreground and at the same time contribute to the lensing signal of a background object, which induces gravitational shear-intrinsic ellipticity correlation (GI, \citealp{hirata04}).

Detailed investigations have been performed on the alignment between
haloes \citep{croft00,heavens00,lee00,catelan01,crittenden01,jing02,mackey02,hirata04,bridle07a,schneiderm09}, as well as the alignment of the spin or the shape of a galaxy with its
own dark matter 
halo (e.g. \citealp{pen00,bosch02,okumura08,okumura09,brainerd09}; see also \citealp{schaefer08}). Intrinsic alignments have also been investigated observationally, where non-vanishing II and GI signals have been detected in several surveys \citep{brown02,heymans04,mandelbaum06,hirata07,brainerd09}.

The results of both theoretical studies and observations show large variations, but most are consistent with a contamination of the order $10\,\%$ by both II and GI correlations for future surveys that further divide the galaxy sample into redshift slices (cosmic shear tomography). Hence, the control of intrinsic alignments in cosmic shear studies is crucial to obtain unbiased results on cosmological parameters. Accurate models would
solve the problem, 
but progress is hampered due to the dependence of intrinsic alignments on the intricacies of galaxy formation and evolution within their dark matter environment. Currently, the level of models is crude, and partly only motivated phenomenologically (see e.g. \citealp{schneiderm09} for recent progress).

The II contamination
can be controlled relatively easily by excluding close pairs of galaxies from the analysis \citep{king02,king03,heymans03,takada04b}. \citet{joachimi08b,joachimi09} introduced a nulling technique which transforms the cosmic shear data vector and discards all entries of the transformed data set that are potentially contaminated by the GI signal. While this approach only relies on the well-known redshift dependence of gravitational lensing, \citet{king05} projects out the GI term by making use of template functions. Furthermore the work by \citet{mandelbaum06} and \citet{hirata07} suggests that the intrinsic alignment may be dominated by luminous red galaxies which could be eliminated from the cosmic shear catalogues. All these removal techniques require excellent redshift information, and still they
can cause a significant 
reduction in the constraints on cosmology.

Deep imaging surveys
not only provide information about the shape of galaxies, but allow in addition for a measurement of galaxy number densities, as well as cross-correlations between shape and number density information. This substantial extension of the set of observables increases the cosmological information to be extracted and, more importantly, enables one to internally calibrate systematic effects \citep{hu03,bernstein08}. By adding galaxy number density information one adds signals that are capable of pinning down the functional form of intrinsic alignments, but one also introduces as another systematic, the galaxy bias, which quantifies the lack of knowledge about how galaxies, i.e. the visible baryonic matter, follow the underlying dark matter distribution.

It is the scope of this work to elucidate the performance of a joint analysis of galaxy shape and number density information as regards the ability to constrain cosmological parameter in presence of general and flexible models of intrinsic alignments and galaxy bias. In doing so we incorporate several cosmological signals which have been considered before as promising probes of cosmology themselves, including galaxy clustering
from photometric redshift surveys 
\citep[][]{blake05,dolney06,zhan06,blake07,padmanabhan07} 
galaxy-galaxy lensing
\citep[e.g.][]{schneider97,guzik01,guzik02,seljak02,seljak05,yoo06,johnston07,cacciato08} 
and lensing magnification
\citep[][]{broadhurst95,zhang05,zhang06,vwaer09}. 
We follow the ansatz outlined in \citet{bernstein08} and extend the investigation by \citet{bridle07} who considered the residual information content in galaxy shape correlations after marginalising over the parameters of two log-linear grid models representing the II and GI terms. We quantify the cross-calibration properties of the joint set of observables and determine the requirements on cosmological surveys to efficiently apply this joint approach.

This paper is organised as follows: In Sect.$\,$\ref{sec:method} we give an overview on the two-point correlations that form part of the galaxy shape and number density observables, and we derive their explicit form. Two appendices provide further details. Section \ref{sec:modelling} demonstrates how we model the different signals and their dependence on cosmology. We introduce a general grid parametrisation for the intrinsic alignments and the galaxy bias. Furthermore we summarise our Fisher matrix formalism and the figures of merit we employ. In Sect.$\,$\ref{sec:results} we present our results on the dependence of the parameter constraints on the freedom in the model of intrinsic alignments and galaxy bias, the characteristics of the redshift distributions, and the priors on the different sets of nuisance parameters. Finally, in Sect.$\,$\ref{sec:conclusions} we summarise our findings and conclude.

\section{Two-point correlations from cosmological surveys}
\label{sec:method}

Cosmological imaging surveys observe the angular positions and the projected shapes of huge numbers of galaxies over increasingly large areas on the sky. In addition, by means of multi-colour photometry, it is possible to perform a tomographic analysis, i.e. obtain coarse information about the line-of-sight dimension in terms of photometric redshifts (photo-z). From the galaxy shapes in a given region of space, one can infer the ellipticity
\eq{
\label{eq:sumkappa}
\epsilon^{(i)}(\vek{\theta}) = \gamma_{\rm G}^{(i)}(\vek{\theta}) + \gamma_{\rm I}^{(i)}(\vek{\theta}) + \epsilon^{(i)}_{\rm rnd}(\vek{\theta})\;,
}
where the superscript in parentheses assigns a photo-z bin $i$. The observed ellipticity $\epsilon$ has contributions from the gravitational shear $\gamma_{\rm G}$ and an intrinsic shear $\gamma_{\rm I}$, which is caused by the alignment of a galaxy in its surrounding gravitational field. Moreover, $\epsilon$ is assumed to have an uncorrelated component $\epsilon_{\rm rnd}$, which accounts for the purely random part of the intrinsic orientations and shapes of galaxies. Note that (\ref{eq:sumkappa}) is only valid if the gravitational shear is weak, see e.g. \citet{seitz97,bartelmann01}
and for certain definitions of ellipticity. 

Likewise, the positions of galaxies can be used to construct an estimate of the number density contrast
\eq{
\label{eq:sumg}
n^{(i)}(\vek{\theta}) = n^{(i)}_{\rm m}(\vek{\theta}) + n_{\rm g}^{(i)}(\vek{\theta}) + n^{(i)}_{\rm rnd}(\vek{\theta})\;,
}
which is determined by the intrinsic number density contrast of galaxies $n_{\rm g}$ and the alteration of galaxy counts due to lensing magnification $n_{\rm m}$. An uncorrelated shot noise contribution is added via $n_{\rm rnd}$. 
In contrast to $\epsilon^{(i)}(\vek{\theta})$ the number density contrast $n^{(i)}(\vek{\theta})$ can obviously not be estimated from individual galaxies. One can understand $n^{(i)}(\vek{\theta})$ as the ensemble average over a hypothetical, Poisson-distributed random field of which the observed galaxy distribution is one particular representation. The formal relation between the projected number density contrast as used in (\ref{eq:sumg}) and the three-dimensional galaxy number density fluctuations will be provided below, see (\ref{eq:projectiongal}).

As was already noted in \citet{bernstein08}, (\ref{eq:sumkappa}) and (\ref{eq:sumg}) are symmetric in the sense that they both contain an intrinsic contribution and a term caused by gravitational lensing effects. Under usual circumstances the correlated part of the ellipticity is dominated by the gravitational shear, whereas the largest term in (\ref{eq:sumg}) is due to the intrinsic number density contrast.

Both ellipticity and number over-density vanish if averaged over sufficiently large scales. Thus, one considers to lowest order two-point statistics of these quantities. Since all real-space two-point measures are related to the power spectrum \citep[see e.g.][]{kaiser92}, we can work in terms of power spectra without loss of generality, which is desirable in particular due to a simpler structure of the signal covariances in Fourier space. Denoting the Fourier transform by a tilde, the power spectrum
$C^{(ij)}_{\rm ab}(\ell)$ between redshift bins $i$ and $j$ 
can then be defined by
\eq{
\label{eq:defpowerspectrum}
\ba{\tilde{x}^{(i)}_{\rm a}(\vek{\ell})\, \tilde{x}^{(j)}_{\rm b}(\vek{\ell'})} = (2\pi)^2 ~\delta^{(2)}_{\rm D}(\vek{\ell} - \vek{\ell'})\; C^{(ij)}_{\rm ab}(\ell)\;,
}
where $\delta^{(2)}_{\rm D}$ is the two-dimensional Dirac delta-distribution, and where $\vek{\ell}$ denotes the angular frequency, the Fourier variable on the sky. The measures $x_{\rm a}$ and $x_{\rm b}$ can correspond to any of the set $\bc{\gamma_{\rm G},\gamma_{\rm I},n_{\rm g},n_{\rm m}}$. The random contributions in (\ref{eq:sumkappa}) and (\ref{eq:sumg}) are not correlated with any of the other measures and only yield a contribution to the noise, see Sect.$\,$\ref{sec:constraints}.

Inserting (\ref{eq:sumkappa}) and (\ref{eq:sumg}) into (\ref{eq:defpowerspectrum}), one obtains the complete set of tomographic two-point
observables 
which are available from shape and number density information
\eqa{
\label{eq:observedps}
C^{(ij)}_{\epsilon \epsilon}(\ell) &=& C^{(ij)}_{\rm GG}(\ell) + C^{(ij)}_{\rm IG}(\ell) + C^{(ji)}_{\rm IG}(\ell) + C^{(ij)}_{\rm II}(\ell)\;\\
\label{eq:psnn}
C^{(ij)}_{n n}(\ell) &=& C^{(ij)}_{\rm gg}(\ell) + C^{(ij)}_{\rm gm}(\ell) + C^{(ji)}_{\rm gm}(\ell) + C^{(ij)}_{\rm mm}(\ell)\;\\
\label{eq:psneps}
C^{(ij)}_{n \epsilon}(\ell) &=& C^{(ij)}_{\rm gG}(\ell) + C^{(ij)}_{\rm gI}(\ell) + C^{(ij)}_{\rm mG}(\ell)+ C^{(ij)}_{\rm mI}(\ell)\;,
}
see \citet{bernstein08}. We name signals stemming from galaxy shape information by capital letters (\lq G\rq\ for gravitational shear, \lq I\rq\ for intrinsic shear) and signals related to galaxy number densities by small letters (\lq g\rq\ for intrinsic number density fluctuations, \lq m\rq\ for lensing magnification). An overview of the nomenclature of the correlations in (\ref{eq:observedps}) to (\ref{eq:psneps}) is provided in Table \ref{tab:correlations}. Note that (\ref{eq:observedps}) and (\ref{eq:psnn}) are symmetric with respect to their photo-z bin arguments. Hence, if $N_{\rm zbin}$ denotes the number of available photo-z bins, one has $N_{\rm zbin}(N_{\rm zbin}+1)/2$ observables for every considered angular frequency. In contrast, one can exploit $N_{\rm zbin}^2$ ellipticity-number density cross-correlation power spectra (\ref{eq:psneps}) per $\ell$.

The set of observables in (\ref{eq:observedps}) is the one that 
cosmic shear
analyses 
are based on. The shear correlation signal (GG) is a clean probe of the underlying matter power spectrum and is thus powerful in constraining cosmological parameters \citep[e.g.][]{hu99}. However, shape measurements incorporate further terms stemming from correlations of intrinsic ellipticities (II) and shear-intrinsic cross-correlations
(IG, or equivalently GI) 
whose contribution can be substantial, but is to date poorly known
\citep{hirata04}. 
These terms exist because the shapes and orientations of galaxies are influenced via the tidal gravitational fields of the matter structures in their surrounding, which firstly induce correlations between neighbouring galaxies, and secondly cause correlations by determining the intrinsic shape of a foreground object and adding to the shear signal of a background galaxy.

Intrinsic galaxy clustering (gg) add a strong signal to the correlations of galaxy number densities (\ref{eq:psnn}), but its use to obtain
cosmological 
parameter estimates is limited due
by 
poor knowledge
of 
the galaxy bias \citep[e.g.][]{lahav04}.
Gravitational lensing modifies the flux of objects and thus reduces or increases number counts of galaxies above a certain limiting magnitude,
depending 
on the form of the galaxy luminosity function close to the limiting magnitude. 
This produces 
magnification correlations (mm) and intrinsic number density-magnification cross-correlations (gm).
The gm correlations
occur when a 
foreground mass overdensity (underdensity)
contains 
an overdensity (underdensity) of galaxies and (de)-magnifies background objects along the same line of sight
causing an apparent over- or under-density of galaxies at higher redshift. 

Cross-correlations between galaxy number densities and ellipticities (\ref{eq:psneps}) contain contributions from cross terms between
intrinsic 
clustering and shear (gG),
intrinsic 
clustering and intrinsic shear (gI), magnification and shear 
(mG), and magnification and intrinsic shear (mI). For instance, one expects to find gI and gG signals when a mass structure leads to an overdensity in the local galaxy distribution and influences the intrinsic shape of galaxies at the same redshift or contributes to the shear of background objects.
The latter is the usual galaxy-galaxy lensing signal. 
Because a foreground 
overdensity can in addition enhance galaxy counts due to lensing magnification, the mG and mI signals will also be non-vanishing. The form of all these correlations will be further discussed in Sect.$\,$\ref{sec:modelling}.

All non-random terms in (\ref{eq:sumkappa}) and (\ref{eq:sumg}), given for a photometric redshift bin $i$, can be related to a source term ${\cal S}$, which is a function of spatial coordinates, i.e.
\eq{
\label{eq:generalprojection}
x_{\rm a}^{(i)}(\vek{\theta}) = \int^{\chi_{\rm hor}}_0 \dd \chi\; w^{(i)}(\chi)\; {\cal S}_{\rm a} \br{ f_{\rm K}(\chi) \vek{\theta}, \chi }\;,
}
where we defined a weight function $w$ that depends on the photo-z bin $i$ \citep[for a similar approach see][]{hu03}. Here, $\chi$ denotes comoving distance, and $f_{\rm K}(\chi)$ is the comoving angular diameter distance. If (\ref{eq:generalprojection}) holds for two quantities $x_{\rm a}^{(i)}$ and $x_{\rm b}^{(j)}$, their projected power spectrum is given by the line-of-sight integral of the three-dimensional source power spectrum $P_{{\cal S}_a {\cal S}_b}$ via Limber's equation in Fourier space \citep{kaiser92},
\eq{
\label{eq:generallimber}
C^{(ij)}_{\rm ab}(\ell) = \int^{\chi_{\rm hor}}_0 \dd \chi\; \frac{w^{(i)}(\chi)\; w^{(j)}(\chi)}{f^2_{\rm K}(\chi)}\; P_{{\cal S}_a {\cal S}_b} \br{\frac{\ell}{f_{\rm K}(\chi)},\chi}\;.
}
By identifying weights and source terms for gravitational and intrinsic shear, as well as intrinsic clustering and magnification, we can derive Limber equations for all power spectra entering (\ref{eq:observedps})-(\ref{eq:psneps}).

To compute the equivalent of (\ref{eq:generalprojection}) for the cosmic shear case, we first note that in Fourier space the shear and the convergence are related by the simple equation $\tilde{\kappa}_{\rm G}(\vek{\ell})=\tilde{\gamma}_{\rm G}(\vek{\ell}) \expo{-2\ic \varphi_\ell}$, where $\varphi_\ell$ is the polar angle of $\vek{\ell}$. As a consequence, the power spectra of shear and convergence are identical. Therefore, we can equivalently use the convergence $\kappa^{(i)}(\vek{\theta})$ as the cosmic shear observable. It is related to the three-dimensional matter density contrast $\delta$ via
\eq{
\label{eq:projectionlensing}
\kappa_{\rm G}^{(i)}(\vek{\theta}) = \int^{\chi_{\rm hor}}_0 \dd \chi\; q^{(i)}(\chi)\; \delta \br{ f_{\rm K}(\chi) \vek{\theta}, \chi }\;,
}
where the weight is given by
\eq{
\label{eq:weightlensing}
q^{(i)}(\chi) = \frac{3 H_0^2 \Omega_{\rm m}}{2\, c^2} \frac{f_{\rm K}(\chi)}{a(\chi)} \int_{\chi}^{\chi_{\rm hor}} \dd \chi'\; p^{(i)}(\chi')\; \frac{f_{\rm K}(\chi' - \chi)}{f_{\rm K}(\chi')}\;,
}
see \citet{bartelmann01,schneider06} for details. Here $a$ denotes the scale factor and $p^{(i)}(\chi)$ the comoving distance probability distribution of those galaxies
in bin $i$ 
for which shape information is available.

Analogously to the lensing case, one can define a convergence of the intrinsic shear field $\kappa_{\rm I}^{(i)}(\vek{\theta})$, which is directly related to the intrinsic shear via $\tilde{\kappa}_{\rm I}(\vek{\ell})=\tilde{\gamma}_{\rm I}(\vek{\ell}) \expo{-2\ic \varphi_\ell}$. This intrinsic convergence is a projection of the three-dimensional intrinsic shear field $\bar{\kappa}_{\rm I}$, which can be written as
\eq{
\label{eq:projectionia}
\kappa_{\rm I}^{(i)}(\vek{\theta}) = \int^{\chi_{\rm hor}}_0 \dd \chi\; p^{(i)}(\chi)\;  \bar{\kappa}_{\rm I} \br{ f_{\rm K}(\chi) \vek{\theta}, \chi }\;,
}
see e.g. \citet{hirata04} for the analogous expression in terms of intrinsic shear. Here we have assumed that the intrinsic shear field is -- like the gravitational shear field -- curl-free to good approximation. This holds for instance for the linear alignment model developed in \citet{hirata04}. Then 
$\bar{\kappa}_{\rm I}$ corresponds to the Fourier transform of $\bar{\gamma}^{\rm I}_{\rm E}(\vek{k})$ as defined in \citet{schneiderm09}.

Likewise, angular galaxy number density fluctuations $n_{\rm g}^{(i)}(\vek{\theta})$ are given by the line-of-sight projection of three-dimensional number density fluctuations $\delta_{\rm g}$ as \citep[e.g.][]{hu03}
\eq{
\label{eq:projectiongal}
n_{\rm g}^{(i)}(\vek{\theta}) = \int^{\chi_{\rm hor}}_0 \dd \chi\; p^{(i)}(\chi)\;  \delta_{\rm g} \br{ f_{\rm K}(\chi) \vek{\theta}, \chi }\;.
}
It is important to note that $p^{(i)}(\chi)$ is the same as in (\ref{eq:weightlensing}) and (\ref{eq:projectionia}), i.e. the number counts are restricted to those galaxies with shape measurements, which require a higher signal-to-noise than the position determination. In principle, number density information could be obtained for a larger number of galaxies, in particular fainter ones. But, to determine the contribution to number density correlations by magnification, it is necessary to measure the slope of the luminosity function $\alpha^{(i)}$ at the faint end of the used galaxy distribution.
We will detail the exact definition and the determination of $\alpha^{(i)}$ in Sect.$\,$\ref{sec:alphalum}. 
Since it is desirable to extract the values of the $\alpha^{(i)}$ internally from the survey, one needs to be able to measure fluxes down to values slightly below the magnitude limit of the galaxies included in $p^{(i)}(\chi)$.
Moreover galaxy number density measurements may require photometric redshifts which are of the same or better quality than for cosmic shear studies, limiting the number of faint usable galaxies. 
Hence, we argue that the choice of identical distance probability distributions for both shape and number density signals is a fair assumption.
We add the warning that one may have to account for selection biases, for instance if one investigates cosmic shear and magnification effects with the same galaxy sample \citep[e.g.][]{schmidt09,krause09}. 

We write the contribution of magnification effects to the number density measurement as
\eqa{
\label{eq:projectionmag}
n^{(i)}_{\rm m}(\vek{\theta}) &=&  2\, (\alpha^{(i)}-1) \int^{\chi_{\rm hor}}_0 \dd \chi\; q^{(i)}(\chi)\; \delta \br{ f_{\rm K}(\chi) \vek{\theta}, \chi }\\ \nn
&=& 2\, (\alpha^{(i)}-1)\; \kappa_{\rm G}^{(i)}(\vek{\theta})\;.
}
As several assumptions enter this equation, in particular concerning the treatment of the slope $\alpha^{(i)}$, we provide a detailed derivation of (\ref{eq:projectionmag}) in Appendix \ref{app:magnification}. Comparing the projection equations (\ref{eq:projectionlensing}), (\ref{eq:projectionia}), (\ref{eq:projectiongal}), and (\ref{eq:projectionmag}) to the general form (\ref{eq:generalprojection}), one can derive all possible cross- and auto-power spectra in the form of the general Limber equation (\ref{eq:generallimber}). For completeness we have given these Limber equations in Appendix \ref{app:limber}.

\begin{table}[t]
\caption{Overview on the two-point correlations considered in this work.}
\centering
\begin{tabular}[t]{lcc}
\hline\hline
measured correlation & 2D PS & 3D PS\\
\hline
shear                            & $C_{\rm GG}$           & $P_{\delta \delta}$\\
intrinsic-shear                         & $C_{\rm IG}$           & $P_{\delta {\rm I}}$\\
intrinsic                               & $C_{\rm II}$           & $P_{\rm II}$\\
galaxy clustering                       & $C_{\rm gg}$           & $P_{\rm gg}$\\
clustering-magnification                    & $C_{\rm gm}$           & $P_{{\rm g} \delta}$\\
magnification                           & $C_{\rm mm}$           & $P_{\delta \delta}$\\
clustering-shear                        & $C_{\rm gG}$           & $P_{{\rm g} \delta}$\\
clustering-intrinsic                    & $C_{\rm gI}$           & $P_{\rm gI}$\\
magnification-shear                     & $C_{\rm mG}$           & $P_{\delta \delta}$\\
magnification-intrinsic                 & $C_{\rm mI}$           & $P_{\delta {\rm I}}$\\
galaxy ellipticity (observable)         & $C_{\epsilon \epsilon}$ &  \\
galaxy number density (observable)      & $C_{n n}$              &  \\
number density-ellipticity (observable) & $C_{n \epsilon}$       &  \\
\hline
\end{tabular}
\tablefoot{Listed are the symbols used for the two-dimensional projected power spectra and the underlying three-dimensional power spectra.}
\label{tab:correlations}
\end{table}

Good 
models of the three-dimensional source power spectra in the Limber equations (see also
the right hand column of 
Table \ref{tab:correlations}) are unknown except for the
non-linear theory 
matter power spectrum $P_{\delta \delta}$. The distribution of galaxies is expected to follow the distribution of dark matter, so that the galaxy clustering power spectra should be related to $P_{\delta \delta}$. However, to date it is unknown how much the galaxy clustering deviates from dark matter clustering, in particular on small scales. This is usually expressed in terms of the galaxy bias $b_{\rm g}$, which is a function of both angular
scale 
$k$ and
redshift or line-of-sight 
distance $\chi$. Hence, one can write
\eqa{
\label{eq:parameterizePg}
P_{\rm gg}(k,\chi) &=& b_{\rm g}^2(k,\chi)\; P_{\delta \delta}(k,\chi)\;\\ \nn
P_{{\rm g} \delta}(k,\chi) &=& b_{\rm g}(k,\chi)\; r_{\rm g}(k,\chi)\; P_{\delta \delta}(k,\chi)\;,
}
where to describe the cross-correlation between matter and galaxy clustering, we introduced a correlation coefficient $r_{\rm g}$ in the second equality.

The intrinsic alignment power spectra depend on the intricacies of galaxy formation and evolution within their dark matter environment. Again, precise models of the intrinsic alignment have to rely on baryonic physics and are currently not available. For symmetry reasons we parametrise our lack of knowledge about the intrinsic alignment power spectra similarly to the galaxy bias as
\eqa{
\label{eq:parameterizePI}
P_{\rm II}(k,\chi) &=& b_{\rm I}^2(k,\chi)\; P_{\delta \delta}(k,\chi)\;\\ \nn
P_{\delta {\rm I}}(k,\chi) &=& b_{\rm I}(k,\chi)\; r_{\rm I}(k,\chi)\; P_{\delta \delta}(k,\chi)\;
}
with the intrinsic alignment bias $b_{\rm I}$ and correlation coefficient $r_{\rm I}$ \citep[following][]{bernstein08}. Although the power spectrum $P_{\rm gI}$ could in principle contain a third, independent correlation coefficient, we assume that it is sufficient to write
\eq{
\label{eq:parameterizePgI}
P_{\rm gI}(k,\chi) = b_{\rm I}(k,\chi)\; r_{\rm I}(k,\chi)\; b_{\rm g}(k,\chi)\; r_{\rm g}(k,\chi)\; P_{\delta \delta}(k,\chi)\;,
}
i.e. we hypothesise that correlations between intrinsic number density fluctuations and intrinsic alignments can entirely be traced back to the effects of the intrinsic alignment bias and the galaxy bias. 
This is a strong assumption since instead of introducing a fifth completely unconstrained bias term, (\ref{eq:parameterizePgI}) establishes a link between the galaxy bias and intrinsic alignment biases. 

Our equation (\ref{eq:parameterizePgI}) is effectively included within the last term in curly brackets of \citet{bernstein08}, eq. (19), and we have effectively set $s^{g \kappa}=0$ in \citet{bernstein08}, eq. (35). \citet{bernstein08} fixes his $s^{g \kappa}$ to be a single unknown scalar across the survey, stating that we expect this type of cross-correlation to have a  minimal effect on cosmological constraints. It would be interesting to check this by comparing results in which  $s^{g \kappa}$ is allowed to vary with those in which  $s^{g \kappa}=0$. However this is beyond the scope of this paper.
 
We note that the unknown quantity in question is the cross-correlation between the intrinsic alignment field and the galaxy position field ($r^{g\kappa}$ in the \citealp{bernstein08} notation), which is precisely the correlation measured in the observational constraints papers \citep[e.g.][]{mandelbaum06,hirata07,mandelbaum09}. Our ansatz states that this is simply related to the cross-correlation between the intrinsic alignment field and the mass, and the cross-correlation between the galaxy position field and the mass ($r^{g\kappa}= r^g r^{\kappa}$ in the \citealp{bernstein08} notation). These latter two quantities are much harder to measure because we do not know the mass field well. Indeed gravitational lensing and the observations we discuss in this paper are likely the best probe of these quantities. Ultimately these data could be used to constrain this additional freedom.

To shed further light on this question, we now discuss a simple toy model to illustrate the nature of our assumption. In the simplest case the galaxy distribution may trace the mass distribution, and the intrinsic shear field is the gradient of the mass distribution. In a toy universe we could displace the position mass field on the sky relative to the other two fields which would render the correlation coefficient $r_I$ (or $r^{\kappa}$ in the notation of \citealp{bernstein08}) zero and also the correlation $r_g=0$ (for large displacements), leaving a strong correlation between the intrinsic shear field and the galaxy position field (unity $r^{g\kappa}$ in the \citealp{bernstein08} notation, thus  $s^{g \kappa}=1$).

Similar more physical arguments could be made by considering stochasticity instead of displacements. However, we might expect the physical origin of both the galaxy position field and the intrinsic shear field to both lie in the mass field, since most of the interactions are mediated by gravity. Therefore it seems reasonable to expect there to be no additional interplay between the intrinsic shear field and galaxy position fields that would produce further cross-correlations (non-zero  $s^{g \kappa}$ in the \citealp{bernstein08} notation).

We will not limit the values of the correlation coefficients to the interval $\bb{-1;+1}$. It is formally possible that $|r|>1$ if our assumption about the statistics of the galaxy distribution, usually taken to be Poissonian, is incorrect \citep{bernstein08}. Treating the correlation coefficients as completely free parameters, our choice of parametrisation in (\ref{eq:parameterizePg}) and (\ref{eq:parameterizePI}) is equivalent to modelling $P_{\rm gg}$ and $P_{{\rm g} \delta}$, or likewise $P_{\rm II}$ and $P_{\delta {\rm I}}$, independently.

We insert the parametrisations (\ref{eq:parameterizePg}), (\ref{eq:parameterizePI}), and (\ref{eq:parameterizePgI}) into the set of Limber equations and can this way relate all power spectra entering (\ref{eq:observedps})-(\ref{eq:psneps}) to the three-dimensional matter power spectrum:
\eqa{
\label{eq:signals_first}
C^{(ij)}_{\rm GG}(\ell) &=& \int^{\chi_{\rm hor}}_0 \dd \chi\; \frac{q^{(i)}(\chi)\; q^{(j)}(\chi)}{f^2_{\rm K}(\chi)}\; P_{\delta \delta} \br{\frac{\ell}{f_{\rm K}(\chi)},\chi}\;\\
C^{(ij)}_{\rm IG}(\ell) &=& \int^{\chi_{\rm hor}}_0 \dd \chi\; \frac{p^{(i)}(\chi)\; q^{(j)}(\chi)}{f^2_{\rm K}(\chi)}\;\\ \nn
&& \times\;  b_{\rm I}\br{\frac{\ell}{f_{\rm K}(\chi)},\chi}\; r_{\rm I}\br{\frac{\ell}{f_{\rm K}(\chi)},\chi}\; P_{\delta \delta} \br{\frac{\ell}{f_{\rm K}(\chi)},\chi}\;\\
\label{eq:signals_II}
C^{(ij)}_{\rm II}(\ell) &=& \int^{\chi_{\rm hor}}_0 \dd \chi\; \frac{p^{(i)}(\chi)\; p^{(j)}(\chi)}{f^2_{\rm K}(\chi)}\;\\ \nn
&& \times\;   b_{\rm I}^2\br{\frac{\ell}{f_{\rm K}(\chi)},\chi}\; P_{\delta \delta} \br{\frac{\ell}{f_{\rm K}(\chi)},\chi}\;\\
C^{(ij)}_{\rm gg}(\ell) &=& \int^{\chi_{\rm hor}}_0 \dd \chi\; \frac{p^{(i)}(\chi)\; p^{(j)}(\chi)}{f^2_{\rm K}(\chi)}\;\\ \nn
&& \times\; b_{\rm g}^2\br{\frac{\ell}{f_{\rm K}(\chi)},\chi}\; P_{\delta \delta} \br{\frac{\ell}{f_{\rm K}(\chi)},\chi}\;\\
C^{(ij)}_{\rm gm}(\ell) &=& 2\,(\alpha^{(j)}-1)\; C^{(ij)}_{\rm gG}(\ell)\;\\
C^{(ij)}_{\rm mm}(\ell) &=& 4\,(\alpha^{(i)}-1)\; (\alpha^{(j)}-1)\; C^{(ij)}_{\rm GG}(\ell)\;\\
C^{(ij)}_{\rm gG}(\ell) &=& \int^{\chi_{\rm hor}}_0 \dd \chi\; \frac{p^{(i)}(\chi)\; q^{(j)}(\chi)}{f^2_{\rm K}(\chi)}\;\\ \nn
&& \times\; b_{\rm g}\br{\frac{\ell}{f_{\rm K}(\chi)},\chi}\; r_{\rm g}\br{\frac{\ell}{f_{\rm K}(\chi)},\chi}\; P_{\delta \delta} \br{\frac{\ell}{f_{\rm K}(\chi)},\chi}\;\\ \nn
C^{(ij)}_{\rm gI}(\ell) &=& \int^{\chi_{\rm hor}}_0 \dd \chi\; \frac{p^{(i)}(\chi)\; p^{(j)}(\chi)}{f^2_{\rm K}(\chi)}\; b_{\rm g}\br{\frac{\ell}{f_{\rm K}(\chi)},\chi}\; r_{\rm g}\br{\frac{\ell}{f_{\rm K}(\chi)},\chi}\\
&& \times\; b_{\rm I}\br{\frac{\ell}{f_{\rm K}(\chi)},\chi}\; r_{\rm I}\br{\frac{\ell}{f_{\rm K}(\chi)},\chi}\; P_{\delta \delta} \br{\frac{\ell}{f_{\rm K}(\chi)},\chi}\;\\
C^{(ij)}_{\rm mG}(\ell) &=& 2\,(\alpha^{(i)}-1)\; C^{(ij)}_{\rm GG}(\ell)\;\\
\label{eq:signals_last}
C^{(ij)}_{\rm mI}(\ell) &=& 2\,(\alpha^{(i)}-1)\; C^{(ji)}_{\rm IG}(\ell)\;.
}
The matter power spectrum and the distances $f_{\rm K}(\chi)$, which also enter $q^{(i)}(\chi)$, depend on cosmology and can therefore be exploited to constrain cosmological parameters. While distances and $P_{\delta \delta}$ are well known from theory, the probability distribution of galaxies $p^{(i)}(\chi)$ has to be measured
by using additional spectroscopic redshift information
\citep[e.g.][]{huterer06,ma05,abdalla07,bridle07,bernstein09} 
with a certain level of uncertainty.
It may also be possible to infer some additional information from the cosmic shear data itself~\citep{newman08,mschneider06,zhang09}. 
The same holds for the slopes of the luminosity function $\alpha^{(i)}$, which can be determined from the survey by studying the flux of galaxies close to the magnitude limit. The least known quantities in the equations above are the bias terms $\bc{b_{\rm g},b_{\rm I},r_{\rm g},r_{\rm I}}$, for which we will thus introduce a very general parametrisation in Sect.$\,$\ref{sec:biasterms}.

\section{Modelling}
\label{sec:modelling}

In this section we detail the modelling of the terms entering
(\ref{eq:signals_first}) to
(\ref{eq:signals_last}). We specify how we parametrise the uncertainty in the galaxy
redshift 
distributions, the slope of the luminosity function, and the bias terms. Moreover we describe our Fisher matrix approach and the way we infer the resulting errors on cosmological parameters.

\subsection{Matter power spectrum \& survey characteristics}
\label{sec:psandsurvey}

As the basis for our analysis we compute matter power spectra for a spatially flat CDM universe with fiducial parameters $\Omega_{\rm m}=0.25$, $\Omega_{\rm DE}=0.75$, and $H_0=100\, h_{100}\, {\rm km/s/Mpc}$ with $h_{100}=0.7$. We incorporate a variable dark energy model by parametrising its equation of state, relating pressure $p_{\rm DE}$ to density $\rho_{\rm DE}$, as \citep{chevallier01,linder03}
\eq{
\label{DEeos}
p_{\rm DE}(z) = \br{w_0 + w_a \frac{z}{1+z}}\; \rho_{\rm DE}(z)\; c^2\;,
}
where the $\Lambda$CDM Universe is chosen as the fiducial model, i.e. $w_0=-1$ and $w_a=0$. The dark energy density parameter is then given by integrating eq.$\,$(3) of \citet{linder03},
\eq{
\label{eq:darkenergy}
\Omega_{\rm DE}(z) = \Omega_{\rm DE}\; \exp \bc{ 3 \br{\br{w_0 + w_a + 1} \ln (1+z) - w_a \frac{z}{1+z}}}\;.
}
The primordial power spectrum of matter density fluctuations is 
assumed to be a power law with fiducial 
slope $n_{\rm s}=1$. We employ the
fiducial 
normalisation $\sigma_8=0.8$. 
The transfer function of \citet{eisenstein98} is used without baryonic wiggles, computing the shape parameter with a fiducial value of $\Omega_{\rm b}=0.05$. 

The non-linear corrections to the power spectrum are computed by means of the fit formula by \citet{smith03}. We account for the influence of dark energy on structure growth by modifying the halo model fitting routine of \citet{smith03}
following the approach of \citet{refregier08}. 
We provide a summary of this modification in Appendix \ref{sec:defitting}.

The survey characteristics
follow the
rough 
specifications of
a Stage IV experiment~\citep{albrecht06} such as 
the ESA Euclid satellite mission.
To compute the noise properties, we assume the maximum extragalactic sky coverage of $A=20000\,{\rm deg}^2$ and a total number density of galaxies $n=35\,{\rm arcmin}^{-2}$. Shape noise is characterised by a total dispersion of intrinsic ellipticities of $\sigma_\epsilon=0.35$.
We refer to this survey as Euclid-like in the remainder of this paper. 

According to \citet{smail94} we assume an overall
number of galaxies per unit redshift, per square arcminute 
\eq{
\label{eq:redshiftdistribution}
n_{\rm tot}(z) = \; \Sigma_0 \; \left(\frac{3z^2}{2\bar{z}^3}\right) \; \exp \bc{ -\br{\frac{z}{\bar{z}}}^\beta}\\ 
}
with the galaxy surface density $\Sigma_0$ and $\beta=1.5$. The 
probability distribution over all the galaxies $p_{\rm tot}(z)$ is proportional to the number density $n_{\rm tot}(z)$. 
We set $\bar{z}=0.64$, which produces a distribution with median redshift $z_{\rm med}=0.9$. The distribution is cut at $z_{\rm max}=3$ and then normalised to unity. For the tomography we define photometric bins by dividing the distribution (\ref{eq:redshiftdistribution}) such that every bin contains the same number of galaxies. This choice is merely for computational convenience and to allow for an easy comparison between results with a different number of bins. As default we will use $N_{\rm zbin}=10$ bins.

To account for photometric redshift errors, we assume that the fraction of catastrophic failures in the assignment of photometric redshifts is negligible, but include the spread of the true redshifts in the bin-wise distributions by writing the conditional probability of obtaining a photometric redshift $z_{\rm ph}$ given the true redshift $z$ as
\eq{
\label{eq:zphgivenzspec}
p(z_{\rm ph}\,|\,z) \propto  \exp \bc{ - \frac{\br{z_{\rm ph} - z}^2}{2\sigma_{\rm ph}^2 \br{1+z}^2 }}\;,
}
where $\sigma_{\rm ph}$ denotes the photometric redshift dispersion. The redshift distribution of an individual photo-z bin $p^{(i)}(z)$ is then obtained by integrating (\ref{eq:zphgivenzspec}) over the bin width and by weighting the result by the overall redshift distribution (\ref{eq:redshiftdistribution}), see \citet{joachimi09} for details.
We 
use
$\sigma_{\rm ph}=0.05$ 
as our default value.
This also follows fiducial Model 1 of~\cite{ma05}.

Since the underlying redshift distributions $p^{(i)}(z)$ are determined by measurement, they are not perfectly known, but introduce further uncertainty into the analysis. A detailed analysis of the dependence of the joint analysis of galaxy shape and number density information on redshift parameters, and also the potential of calibrating these errors internally, will be investigated elsewhere
\citep[e.g.][]{zhang09}. 
\citet{bridle07} have undertaken a more detailed study of the effect of redshift errors in the case of ellipticity correlations only. For the purpose of this work we assume that the value of $\sigma_{\rm ph}$ is unknown, i.e. we use it as a single, global parameter to account for the uncertainty in the redshift distributions. We employ a
wide
Gaussian 
prior on $\sigma_{\rm ph}$ of 10 for reasons of numerical stability.

\subsection{Galaxy luminosity function}
\label{sec:alphalum}

In order to calculate power spectra which include the lensing magnification signal, we need to model the slope of the cumulative galaxy luminosity function at the magnitude limit of the galaxy number density catalogue. In Appendix \ref{app:gallum} we extend observational results for the normalisation and redshift scaling of the galaxy redshift distribution (\ref{eq:redshiftdistribution}) by \citet{blake05} to provide a fitting formula for the luminosity function slope as a function of redshift and survey magnitude limit. 

We use the fit given by (\ref{eq:alphafit1}) and (\ref{eq:alphafit2}) with the parameters listed in Table \ref{tab:lumfit} to compute the slope of the luminosity function at $r_{\rm lim}=24$. The discussion in this work applies to ground-based surveys because the COMBO-17 luminosity functions are calculated for the SDSS r filter as observed from the ground. A space mission to a depth of $r_{\rm lim}=24$ will have a different luminosity function slope, corresponding more closely to a deeper ground based survey, depending on the resolution of the space-based survey. We use results for $r_{\rm lim}=24$ throughout this paper for both ground and space surveys. We note that from Fig.$\,$\ref{fig:plot_alphas}, top panel, the slope of the luminosity function is changed little on increasing the survey depth beyond $r_{\rm lim}=24$. 

The fiducial slope in a photo-z bin $i$ is defined as $\alpha^{(i)} \equiv \alpha(z^{(i)}_{\rm med},r_{\rm lim}=24)$, where $z^{(i)}_{\rm med}$ is the median redshift of bin $i$, see Appendix \ref{app:magnification}. We assume $\alpha^{(i)}$ is also measured from the survey itself, and therefore adds another source of uncertainty to the analysis which we account for by setting $\alpha^{(i)}$ to be free parameters for all $i=1,\,..\,,N_{\rm zbin}$. Again we apply a wide Gaussian prior of 10 on every slope parameter, which does not have a measurable influence on our results, but merely ensures numerical stability.

\subsection{Galaxy and intrinsic alignment bias}
\label{sec:biasterms}

As already outlined in Sect.$\,$\ref{sec:method}, the bias terms $b_X=\bc{b_{\rm g},b_{\rm I},r_{\rm g},r_{\rm I}}$ encoding the galaxy bias and intrinsic alignments are the least accurately known contributions to (\ref{eq:signals_first}) to (\ref{eq:signals_last}). We parametrise each of these terms on a grid in $k$ and $z$, following \citet{bridle07} whose ansatz is in turn
similar to 
the recommendations by the Dark Energy Task Force \citep{albrecht06}
and~\citet{bernstein08}. 
Every bias term is assumed to vary around a fiducial functional form $b^{\rm base}_X$ as
\eq{
\label{eq:biasterms}
b_X(k,\chi) = A_X\; Q_X\br{k,z(\chi)}\; b^{\rm base}_X(k,\chi)\;,
}
where $Q_X\br{k,z(\chi)}$ is an unknown two-dimensional function which comprises the aforementioned grid, and where $A_X$ denotes an additional free overall amplitude. We use $N_K$ bins in $k$ and $N_Z$ bins in redshift for each bias term and linearly interpolate in the logarithms of $Q_X$, $k$, and $1+z$, so that $Q_X$ is given by
\eqa{
\label{eq:Q}
\ln Q_X\br{k,z} &=& K_i(k)\; Z_j(z)\; B^X_{ij} + \bigl[1-K_i(k)\bigr]\; Z_j(z)\; B^X_{(i+1)\,j}\\ \nn
&& \hspace*{-2cm} + K_i(k)\; \bb{1-Z_j(z)}\; B^X_{i\, (j+1)} + \bigl[1-K_i(k)\bigr]\; \bb{1-Z_j(z)}\; B^X_{(i+1)\, (j+1)}
}
for $k_i < k \leq k_{i+1}$ and $z_j < z \leq z_{j+1}$, where we defined 
\eqa{
\label{eq:defKZ}
K_i(k) &\equiv& \frac{\ln(k) - \ln(k_i)}{\ln(k_{i+1}) - \ln(k_i)}\;\\ \nn
Z_j(z) &\equiv& \frac{\ln(1+z) - \ln(1+z_j)}{\ln(1+z_{j+1}) - \ln(1+z_j)}\;.
}
The free parameters are the grid nodes $B^X_{ij}$. Since $Q_X$ is a multiplicative function, one reproduces the base model $b^{\rm base}_X$ if $A_X=1$ and if all $B^X_{ij}$ vanish. The effect of this parametrisation on the observable projected power spectra is illustrated in \citet{bridle07} for the case of intrinsic alignments.

The indices in (\ref{eq:Q}) run from $i=0,\,..\,,N_K+1$ and $0=1,\,..\,,N_Z+1$. We fix all parameters at the edge of the grid by setting the parameters with indices $i,j=0$; $i=N_K+1$ or $j=N_Z+1$ to $B^X_{ij}=0$, so that we have $N_K \times N_Z$ free grid parameters per bias term. We place the lowest and highest grid nodes at the limits of our integration ranges, so $k_0=3.3 \cdot 10^{-7}h\,{\rm Mpc}^{-1}$ and $k_{N_K+1}=3.3 \cdot 10^4h\,{\rm Mpc}^{-1}$ in $k$, and in the redshift dimension $z_0=0$ and $z_{N_Z+1}=19$. The grid nodes, which are free to vary, are log-linearly spaced in a smaller range, respectively. We use $k_1=10^{-3}h\,{\rm Mpc}^{-1}$ and $k_{N_K}=2 h\,{\rm Mpc}^{-1}$, and for the redshift range $z_1=z_0$ and $z_{N_Z}=3$. In the special case of $N_K=1$ we position the only free parameter in the $k$ dimension at the centre between $k_1$ and $k_{N_K}$, and proceed likewise for redshifts.

\begin{table}[t]
\caption{Overview on the total number of nuisance parameters used for different setups.}
\centering
\begin{tabular}[t]{cccc}
\hline\hline
$N_K \times N_Z$ & $\epsilon \epsilon$ & $nn$ & all\\
\hline
 2 $\times$ 2  & 11 & 21 &  31\\
 2 $\times$ 4  & 19 & 29 &  47\\
10 $\times$ 2  & 43 & 53 &  95\\
10 $\times$ 4  & 83 & 93 & 175\\
 3 $\times$ 3  & 21 & 31 &  51\\
 5 $\times$ 5  & 53 & 63 & 115\\
 7 $\times$ 7  &101 &111 & 211\\
\hline
\end{tabular}
\tablefoot{The label $\epsilon \epsilon$ corresponds to using ellipticity correlations only as the observables. Likewise, $nn$ corresponds to using galaxy number density correlations only, and \lq all \rq\ to using all available correlations. For all entries we have assumed ten photometric redshift bins used for the tomography, $N_{\rm zbin}=10$.}
\label{tab:nuispar}
\end{table}

It is important to note that while \citet{bridle07} limit the flexible grid parametrisation to the non-linear regime of the power spectra, we attempt to cover all $k$ ranges which substantially contribute to the observable power spectra. As we fix the grid values on the edges, the overall scaling of the bias terms is not free, so that we use the amplitude $A_X$ as a further varying parameter throughout. To all bias term parameters we add a very wide Gaussian prior of
standard deviation 
50 to ensure numerical stability. Together with the global uncertainty on the redshift distributions, expressed in terms of $\sigma_{\rm ph}$, and the values of the slope of the galaxy luminosity function per photo-z bin, we obtain a large number of nuisance parameters that we determine simultaneously with the cosmological parameters of interest. For later reference, we have summarised the total number of nuisance parameters for different setups in Table \ref{tab:nuispar}. While our parametrisation is fairly general and should capture most of the variability, it is of course possible that the bias terms depend on more parameters than $k$ and $z$. For instance, it is well known that both intrinsic alignments and galaxy bias are a function of galaxy
colour and luminosity 
which could be incorporated into our approach in the future.
For observational constraints on this effect using intrinsic alignments see \cite{mandelbaum06,hirata07} and in galaxy biasing see \cite{mccracken07, swanson08, simon08, cresswell09, wang07} for recent examples. 

\begin{figure*}[ht!!!]
\centering
\includegraphics[scale=0.85]{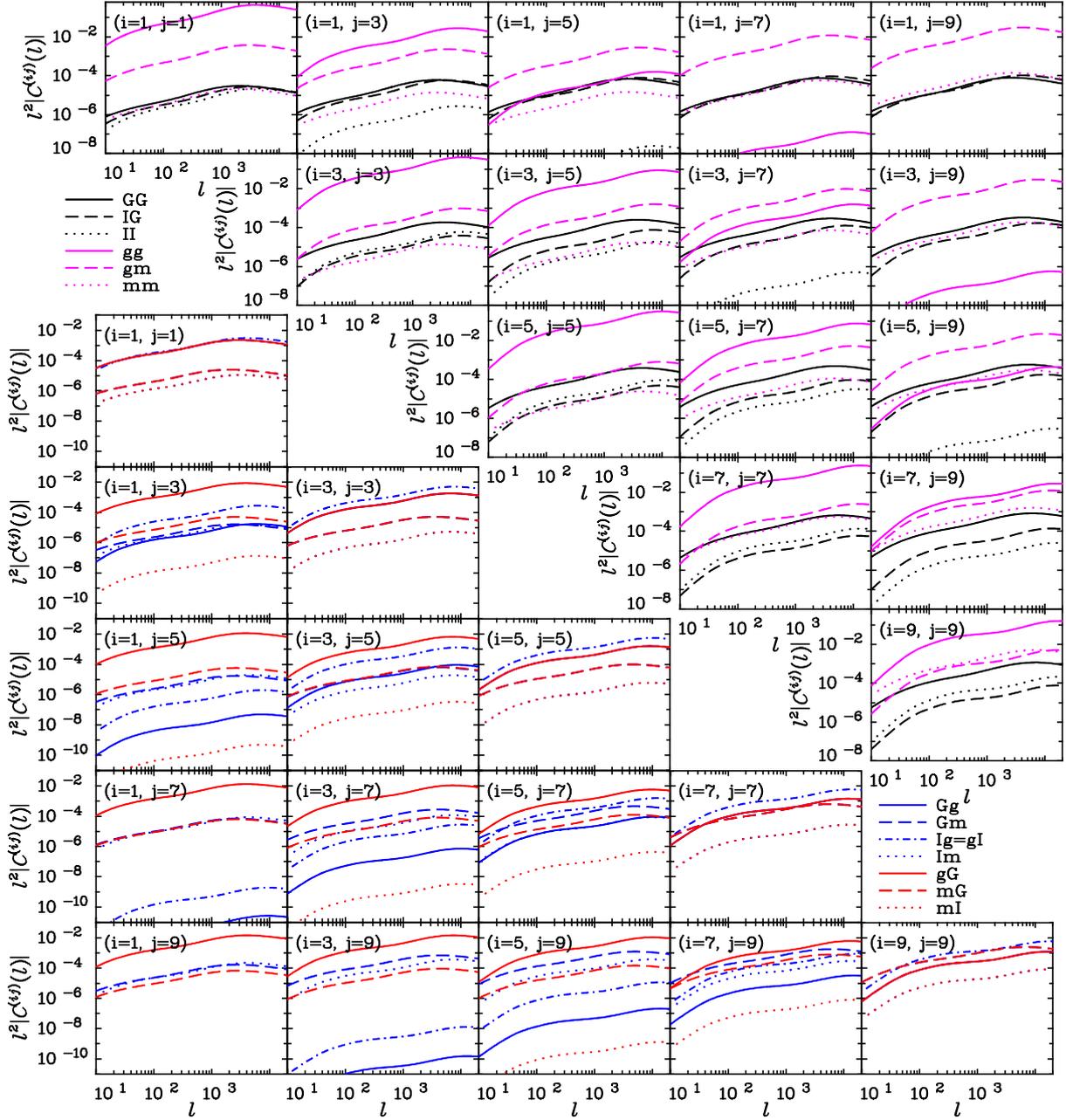}
\caption{Fiducial power spectra for all considered correlations. The upper right panels depict the contributions to $\epsilon\epsilon$ (in black) and $nn$ (in magenta) correlations. 
The lower left panels show the contributions to correlations between number density fluctuations and ellipticity. Since we only show correlations $C_{\alpha \beta}^{(ij)}(\ell)$ with $i \leq j$, we make in this plot a distinction between $n \epsilon$ (in red; number density contribution in the foreground, e.g. gG) and $\epsilon n$ (in blue; number density contribution in the background, e.g. Gg) correlations. 
In each sub-panel a different tomographic redshift bin correlation is shown. For clarity only odd bins are displayed. 
In the upper right panels the usual cosmic shear signal (GG) is shown as a black solid lines; the intrinsic alignment GI term is shown by the black dashed lines; the intrinsic alignment II term is shown by the dotted black line; the usual galaxy clustering signal (gg) is shown by the magenta solid line; the cross correlation between galaxy clustering and lensing magnification (gm) is shown by the magenta dashed line; the lensing magnification correlation functions (mm) are shown by the magenta dotted line. 
In the lower left panels the solid blue line shows the correlation between lensing shear and galaxy clustering (Gg); the blue dashed line shows the correlation between lensing shear and lensing magnification (gm); the blue dot-dashed line shows the correlation between intrinsic alignment and galaxy clustering (Ig or equivalently gI); the red solid line shows the correlation between galaxy clustering and lensing shear (gG), which is equivalent to the blue solid line with redshift bin indices $i$ and $j$ reversed; similarly the red dashed line shows the correlation between lensing magnification and lensing shear (mG), for cases where the magnification occurs at lower redshift than the shear ($i<j$); finally the dotted line shows the correlation between lensing magnification and intrinsic alignment (mI). 
}
\label{fig:plotps}
\end{figure*}

To compute the fiducial models for all the power spectra entering the observables (\ref{eq:observedps}) to (\ref{eq:psneps}), we set $A_X=1$ and all $B^X_{ij}=0$, i.e. they are fully determined by the base models. We set
\eqa{
\label{eq:basemodels}
b_{\rm g}^{\rm base}(k,\chi) &=& 1\;\\ \nn
r_{\rm g}^{\rm base}(k,\chi) &=& 1\;\\ \nn
b_{\rm I}^{\rm base}(k,\chi) &=& -C_1 \frac{\bar{\rho}(z)}{D(z) (1+z)}\;\\ \nn
r_{\rm I}^{\rm base}(k,\chi) &=& 1\;.
}
With the choice for $b_{\rm I}^{\rm base}$ and $r_{\rm I}^{\rm base}$ we reproduce the non-linear modification of the linear alignment model by \citet{bridle07}. Lacking solid physical motivation, it is yet in agreement with current observational evidence (\citealp{mandelbaum06}; see \citealp{bridle07} for a comparison) and the halo model studies by \citet{schneiderm09}. It is based on the linear alignment model \citep{hirata04} which is expected to provide a good description of intrinsic alignments on the largest scales. We assume the galaxy bias to be of order unity for our fiducial model, and set $r_{\rm g}^{\rm base}(k,\chi)=1$. Note that most investigations of galaxy clustering consider much less flexibility in the galaxy bias.

In Fig.$\,$\ref{fig:plotps} we plot the fiducial
angular 
power spectra of all considered signals for different combinations of photo-z bins.
Note that the ellipticity-number density cross-correlations are not symmetric under exchanging the photometric redshift bins. Hence, in this figure we treat n$\epsilon$ and $\epsilon$n correlations, as well as all signals contributing to them, separately, keeping $i \leq j$ for all $C_{\alpha \beta}^{(ij)}(\ell)$. 

The GG signal shows the usual behaviour of moderate increase with increasing redshift of the contributing photo-z-bins. The redshift scaling of the IG term is similar, but peaks when the source galaxies are at high redshift in the background (large $j$), while the galaxies that are intrinsically aligned are at low redshift (small $i$). For the model used here the IG contribution can even surpass the shear signal in this case. Due to the narrow kernel containing $p^{(i)}(\chi)\, p^{(j)}(\chi)$, see (\ref{eq:signals_II}), the II signal is strong in the auto-correlations $i=j$, but drops off quickly as soon as the overlap of the redshift distributions decreases.

Due to the similar kernel, the scaling of the galaxy clustering contribution (gg) resembles the II term, but gg constitutes a much stronger signal. Lensing magnification (mm) adds
the largest fraction of 
the galaxy number correlations at the highest redshifts, showing a slightly stronger redshift scaling than GG. However, the mm term always remains subdominant with respect to signals with a contribution from galaxy clustering; even for widely separated galaxy redshift distributions, say $i=1$ and $j=9$,
and 
the gm cross-term is considerably stronger than mm. Such contributions might be a serious obstacle for probing cosmology with the lensing magnification signal as proposed by
\citet{broadhurst95,zhang05,zhang06,vwaer09}. 
Yet in our approach, where the galaxy bias is taken into account and parametrised, the magnification signal yields a valuable contribution to the galaxy number correlations, which helps constraining the cosmological model.

The signals within the number density-ellipticity cross-correlations are not symmetric when swapping the photo-z bins. When the contribution by number density fluctuations stems from the foreground, the gG signal is strong, in particular if the photo-z bins are far apart in redshift, whereas the \lq Gg\rq\ (in the notation of Fig.$\,$\ref{fig:plotps}) drops off fast if $i < j$ 
because the shear signal of foreground galaxies is not correlated with the clustering of galaxies at much higher redshift. The mG, Gm, and GG signals differ only by the term including the slope of the luminosity function and thus have similar amplitudes. Correlations between intrinsic alignment and magnification (mI and Im) are subdominant throughout, obtaining their largest amplitudes if intrinsic alignments at low redshifts are combined with the magnification signal from galaxies far in the background, i.e. in the figure for Im at $i=1$ and $j=9$. Finally, the symmetric Ig term is the largest contribution for auto-correlations of number density-ellipticity observables, but decreases quickly in the cross terms, again due to the kernel $p^{(i)}(\chi)\, p^{(j)}(\chi)$
since we have assumed the photometric redshift errors are reasonably well behaved, without catastrophic outliers. 
Note that we have plotted the absolute values of the power spectra in Fig.$\,$\ref{fig:plotps} and that the correlations IG, gm, Gm, mG, and gI are negative.

\subsection{Parameter constraints}
\label{sec:constraints}

We determine constraints on our parameters
using 
a Fisher matrix analysis. To account for the errors and correlations of our observables, we compute covariances of the power spectra (\ref{eq:observedps}) to (\ref{eq:psneps}) in the Gaussian approximation, extending the results of \citet{joachimi08}; see also \citet{hu03}. If we denote the difference between estimator and its ensemble average by $\Delta C_{\alpha \beta}^{(ij)}(\ell)$, one can write for the covariance
\eqa{
\label{eq:cov}
\ba{ \Delta C_{\alpha \beta}^{(ij)}(\ell)\; \Delta C_{\gamma \delta}^{(kl)}(\ell') } &=& \delta_{\ell \ell'} \frac{2\, \pi}{A \ell \Delta \ell}\\ \nn
&& \hspace*{-2cm} \times\; \bc{ \bar{C}_{\alpha \gamma}^{(ik)}(\ell)\; \bar{C}_{\beta \delta}^{(jl)}(\ell) + \bar{C}_{\alpha \delta}^{(il)}(\ell)\; \bar{C}_{\beta \gamma}^{(jk)}(\ell) } \equiv {\rm Cov}_{\alpha \beta \gamma \delta}^{(ijkl)}(\ell)\;,
}
where $A$ is the survey size and $\Delta \ell$ the width of the corresponding angular frequency bin. As the Kronecker symbol $\delta_{\ell \ell'}$ indicates, the covariance is diagonal in $\ell$ in the Gaussian limit, which keeps the computation and inversion of (\ref{eq:cov}) tractable. The subscripts $\bc{\alpha, \beta, \gamma, \delta}$ can be either $\epsilon$ or $n$, where $C_{\epsilon n}^{(ij)}(\ell) \equiv C_{n \epsilon}^{(ji)}(\ell)$ holds. To account for the shot and shape noise contributions induced by the random terms in (\ref{eq:sumkappa}) and (\ref{eq:sumg}), we have defined
\eq{
\label{eq:covplusnoise}
\bar{C}_{\alpha \beta}^{(ij)}(\ell) \equiv C_{\alpha \beta}^{(ij)}(\ell) + N_{\alpha \beta}^{(ij)}\;,
}
the second term given by
\eqa{
\label{eq:covnoiseterms}
N_{\alpha \beta}^{(ij)} &=& \delta_{ij}\; \frac{\sigma_\epsilon^2}{2 \bar{n}^{(i)}} ~~~\mbox{for}~ \alpha=\beta=\epsilon\\ \nn
N_{\alpha \beta}^{(ij)} &=& \delta_{ij}\; \frac{1}{\bar{n}^{(i)}} ~~~\mbox{for}~ \alpha=\beta=n\\ \nn
N_{\alpha \beta}^{(ij)} &=& 0 ~~~\mbox{for}~ \alpha \neq \beta\;.
}
Here $\sigma_\epsilon^2$ denotes the total intrinsic ellipticity dispersion, and $\bar{n}^{(i)}$ is the average galaxy number density
per steradian 
in photo-z bin $i$.

Combining the observable power spectra, we compose the total data vector
\eqa{
\label{eq:datavector}
{\cal D}(\ell) &=& \biggl\{ C^{(11)}_{\epsilon \epsilon}(\ell),\,..\,,C^{(N_{\rm zbin} N_{\rm zbin})}_{\epsilon \epsilon}(\ell),C^{(11)}_{n \epsilon}(\ell),\,..\,,\\ \nn
&& \hspace*{1cm} C^{(N_{\rm zbin} N_{\rm zbin})}_{n \epsilon}(\ell),  C^{(11)}_{n n}(\ell),\,..\,,C^{(N_{\rm zbin} N_{\rm zbin})}_{n n}(\ell) \biggr\}^\tau\;
}
for every angular frequency considered. The corresponding covariance, again for every $\ell$, reads
\eq{
\label{eq:covtotal}
{\rm Cov}(\ell) = \br{ \begin{array}{c|c|c}
{\rm Cov}_{\epsilon \epsilon \epsilon \epsilon}^{(ijkl)}(\ell) & {\rm Cov}_{\epsilon \epsilon n \epsilon}^{(ijkl)}(\ell) & {\rm Cov}_{\epsilon \epsilon n n}^{(ijkl)}(\ell)\\
\hline
{\rm Cov}_{n \epsilon \epsilon \epsilon}^{(ijkl)}(\ell) & {\rm Cov}_{n \epsilon n \epsilon}^{(ijkl)}(\ell) & {\rm Cov}_{n \epsilon n n}^{(ijkl)}(\ell)\\
\hline
{\rm Cov}_{n n \epsilon \epsilon}^{(ijkl)}(\ell) & {\rm Cov}_{n n n \epsilon}^{(ijkl)}(\ell) & {\rm Cov}_{n n n n}^{(ijkl)}(\ell)
\end{array} }\;,
}
with the block matrices given by (\ref{eq:cov}). The number of galaxy ellipticity ($\epsilon \epsilon$) and number density ($nn$) observables entering ${\cal D}(\ell)$ is $N_{\rm zbin} \br{N_{\rm zbin}+1}/2$, respectively, while there are $N_{\rm zbin}^2$ ellipticity-number density cross terms ($n \epsilon$), which are not symmetric. In the analysis that follows we will also consider $\epsilon \epsilon$ and $nn$ correlations only. In these cases the covariance (\ref{eq:covtotal}) is reduced accordingly to its upper left or lower right block.

For reasons of computational time the total number of parameters that we can consider is limited to a few hundred. As a consequence the $k$ dependence of the galaxy bias can not be parametrised by more than about ten parameters per redshift grid node. This number might not provide enough freedom in $b_{\rm g}$ (and $r_{\rm g}$) to represent a sufficiently general set of functional forms, which inadvertently may cause strong constraints on cosmological parameters due to the strong signal of galaxy clustering. Hence, we follow existing studies of galaxy clustering by discarding the clustering contribution in the non-linear regime where the signal is largest and the form of the galaxy bias most uncertain.

\citet{rassat08} calculated wave vectors $k_{\rm lin}^{\rm max}$ as a function of redshift at which the three-dimensional power spectrum has to be cut off to avoid number density fluctuations above a certain threshold, used as an indicator for non-linearity. Since we do still have a fairly general parametrisation of the bias terms associated with galaxy bias, we can afford to include the mildly non-linear regime into our analysis. Consulting Fig.$\,$2 of \citet{rassat08}, we 
choose a simple linear parametrisation of the form 
\eq{
\label{eq:kmaxfit}
k_{\rm lin}^{\rm max}(z) \approx 0.132\, z\; h\,{\rm Mpc}^{-1}\;.
}
This relation roughly coincides with the fiducial curve in the figure, producing slightly more conservative cuts at low redshifts. 

We do not cut the three-dimensional power spectrum in $k$-space, but instead exclude projected power spectra above a threshold angular frequency from the Fisher matrix analysis. This maximum angular frequency is computed via
\eq{
\label{eq:lmaxgal}
\ell_{\rm max}^{{\rm g}\;(i)} = k_{\rm lin}^{\rm max}(z^{(i)}_{\rm med})\; f_{\rm k}\br{\chi(z^{(i)}_{\rm med})}\;,
}
where we choose as a characteristic redshift of bin $i$ the median redshift $z^{(i)}_{\rm med}$. Hence, we obtain a cut-off $\ell$ for every photo-z bin.
We choose that 
$\epsilon \epsilon$ correlations are not at all affected by this cut-off
because they are not dominated by terms involving galaxy bias. 
We impose $\ell_{\rm max}^{{\rm g}\;(i)}$ on $n \epsilon$ correlations, where $i$ is the photo-z bin from which the number density signal stems. For observables $C^{(ij)}_{n n}(\ell)$ we use the cut-off calculated for bin $j$. Note that, due to the fast drop-off of the galaxy clustering signal with increasingly different median redshifts of bins $i$ and $j$, the more optimistic choice of bin $j$ over $i$ in the latter case should not influence our results decisively.

\begin{table}[t] 
\caption{Overview on the cuts in angular frequency.}
\centering
\begin{tabular}[t]{cccc}
\hline\hline
$z$-bin & $z_{\rm med}$ & $\ell_{\rm max}^{\rm g}$ & $N^{\rm max}_\ell$\\
\hline
1  & 0.33 & 45   & 13\\
2  & 0.51 & 103  & 20\\
3  & 0.63 & 145  & 23\\
4  & 0.74 & 206  & 26\\
5  & 0.85 & 260  & 28\\
6  & 0.96 & 329  & 30\\
7  & 1.08 & 369  & 31\\
8  & 1.23 & 466  & 33\\
9  & 1.42 & 588  & 35\\
10 & 1.79 & 937  & 39\\
\hline
\end{tabular}
\tablefoot{Results are shown for the default set of parameters with $N_{\rm zbin}=10$ and $\sigma_{\rm ph}=0.05$, using (\ref{eq:kmaxfit}) and (\ref{eq:lmaxgal}). For each photo-z bin the median redshift $z_{\rm med}$, the maximum angular frequency $\ell_{\rm max}^{\rm g}$, and the corresponding number of usable angular frequency bins $N^{\rm max}_\ell$ (out of the total of 50) is given.}
\label{tab:lmax}
\end{table}

To compute the Fisher matrix, we use $N_\ell=50$ angular frequency bins, spaced logarithmically between $\ell_{\rm min}=10$ and $\ell_{\rm max}=3000$, the latter value being a
conservative maximum for future surveys. We assume that the covariance of the power spectra is independent of the cosmological parameters, so that it does not contribute to the constraints. Then the Fisher matrix reads \citep[e.g.][]{tegmark97}
\eq{ 
\label{eq:fishermatrix}
F_{\mu \nu} = \sum_{m,n}^{N_d} \sum_\ell^{N^{\rm max}_\ell(m,n)} \frac{\partial {\cal D}_m (\ell)}{\partial p_\mu}\; {\rm Cov}^{-1}_{mn}(\ell)\; \frac{\partial {\cal D}_n (\ell)}{\partial p_\nu}\;,
}
where $N_d$ is the dimension of ${\cal D} (\ell)$. The first summation in (\ref{eq:fishermatrix}) runs over all $N^{\rm max}_\ell$ usable angular frequency bins. The number of usable bins depends on the type of correlation and is determined by the cut-off angular frequency as described above. For the default setup we have summarised $z^{(i)}_{\rm med}$, $\ell_{\rm max}^{{\rm g}\;(i)}$, and the number of usable bins for every photo-z bin in Table \ref{tab:lmax}. The derivatives in (\ref{eq:fishermatrix}) are taken with respect to the elements of the parameter vector
\eqa{
\label{eq:parametervector}
\vek{p} &=& \biggl\{ \Omega_{\rm m},\sigma_8,h_{100},n_{\rm s},\Omega_{\rm b},w_0,w_a;\\ \nn
&& \hspace*{2cm} \sigma_{\rm ph},\alpha^{(1)},\,..\,,\alpha^{(N_{\rm zbin})},B_{11}^{b_{\rm g}},\;\;..\;\;,B_{N_K\,N_Z}^{r_{\rm I}} \biggr\}^\tau\;.
}
The first seven entries of $\vek{p}$ correspond to the cosmological parameters that we are interested in, while the remaining nuisance parameters account for the uncertainty in the galaxy redshift distribution, the slope of the galaxy luminosity function, the intrinsic alignments, and the galaxy bias, see Table \ref{tab:nuispar}. In summary, we use a maximum of $N_\ell (2 N_{\rm zbin}^2 + N_{\rm zbin})$ observables (actually significantly less due to the $\ell$-cuts of the galaxy number density signals) to measure a total of $4 N_Z N_K + N_{\rm zbin} + 8$ parameters. Note that since we have referred all signals contributing to the observables (\ref{eq:observedps}) to (\ref{eq:psneps}) to the matter power spectrum, they all constrain the set of cosmological parameters; none of them is fixed when calculating (\ref{eq:fishermatrix}).

The minimum variance bound of the error on a parameter $p_\mu$, if determined simultaneously with all other parameters, is given by $\sigma(p_\mu) = \sqrt{(F^{-1})_{\mu\mu}}$. This error provides us with a lower bound on the marginalised $1\,\sigma$-error on $p_\mu$. To assess the statistical power of the survey by means of a single number, we use the figure of merit (FoM) suggested by the Dark Energy Task Force (DETF) Report \citep{albrecht06},
\eq{ 
\label{eq:fomdetf}
{\rm FoM}_{\rm DETF} = \frac{1}{\sqrt{\det \br{F^{-1}}_{w_0 w_a}}}\;,
}
where the subscript \lq$w_0 w_a$\rq\ denotes the $2 \times 2$ sub-matrix of the inverse Fisher matrix that corresponds to the entries belonging to the two dark energy parameters. 
Note that different prefactors for (\ref{eq:fomdetf}) are used in the literature. To allow for direct comparison with \citet{bridle07}, divide our findings for the ${\rm FoM}_{\rm DETF}$ by four. 

While (\ref{eq:fomdetf}) is restricted to the quality of constraints on dark energy, we also seek to consider the errors on all cosmological parameters of interest. We are interested in the total volume of the error ellipsoid in parameter space, which is measured by the determinant of the Fisher matrix. Hence, we define
\eq{ 
\label{eq:fomtot}
{\rm FoM}_{\rm TOT} \equiv\; \ln \br{\, \frac{1}{\det \br{F^{-1}}_{\rm cosm.}} } \;,
}
where only the sub-matrix of the inverse Fisher matrix that corresponds to the seven cosmological parameters under investigation is used in the determinant, as indicated by the subscript. The determinant of the inverse is computed in order to take the effect of marginalising over nuisance parameters into account.

\section{Results}
\label{sec:results}

Based on the Fisher matrix formalism described in the foregoing section, we will now analyse the performance of a cosmological galaxy survey with combined number density and shear information. We are going to investigate the residual information content in the data after marginalising over models of the intrinsic alignments and the galaxy bias with varying degrees of freedom. Furthermore, we will study the dependence of our FoM on the number of photo-z bins and the width of the bin-wise redshift distributions as well as on the priors imposed on the nuisance parameters. The information contained in the individual signals and their susceptibility to the nuisance parameters is also assessed. Throughout this section we use the default survey characteristics and parameter values unless specified otherwise.

\subsection{Dependence on intrinsic alignments and galaxy bias}
\label{sec:resultsnuispar}

The four bias terms $\bc{b_{\rm g},b_{\rm I},r_{\rm g},r_{\rm I}}$ each comprise $N_K \times N_Z +1$ nuisance parameters. The galaxy ellipticity ($\epsilon \epsilon$) power spectra only contain the two intrinsic alignment bias terms, whereas the number density correlations ($nn$) are only affected by the galaxy bias terms. The cross-correlations between ellipticity and number density link those signals which depend on both galaxy bias and intrinsic alignments and thus allow for their internal cross-calibration. An example is the study by \cite{zhang08}
which 
investigates the interrelations between the IG, gI, and gg terms.

\begin{figure*}[ht]
\centering
\includegraphics[scale=.65,angle=270]{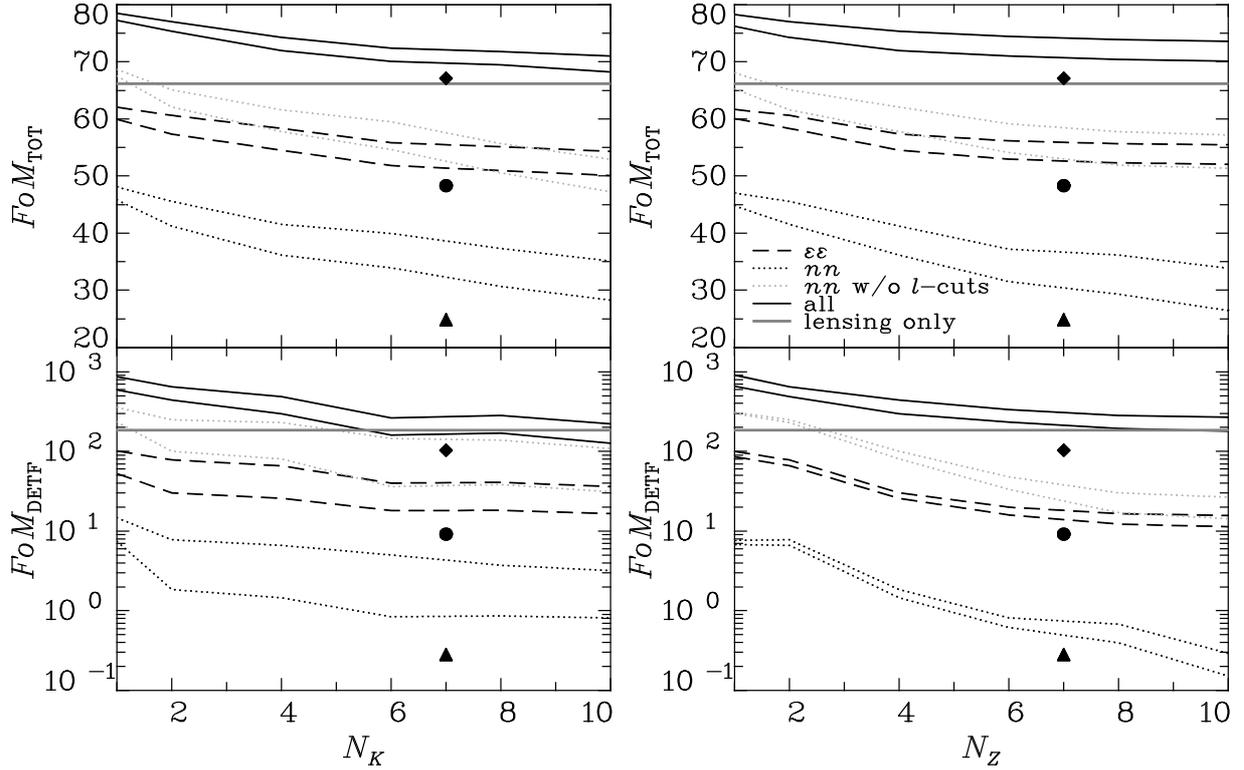}
\caption{
\textit{Left panels}:
Figures of merit 
as a function of the number of
free parameters as a function of 
wave vector
$N_K$ in the bias terms
For each
line type, 
the upper
curve is 
obtained for
a number of free bias parameters as a function of redshift 
$N_Z=2$, the lower
is 
for $N_Z=4$.
\textit{Right panels}: Same as on the left, but as a function of $N_Z$, i.e. the number of redshift parameters in the bias terms. The upper curves for each set correspond now to $N_K=2$ and the lower ones to $N_K=4$, respectively.
\textit{Upper panels}: Figure of merit taking into account the full cosmological parameter space, ${\rm FoM}_{\rm TOT}$, see (\ref{eq:fomtot}). 
\textit{Lower panels}: Dark Energy figure of merit from the Dark Energy Task Force ${\rm FoM}_{\rm DETF}$, see (\ref{eq:fomdetf}). 
Dashed curves correspond to
results using galaxy ellipticity correlations ($\epsilon\epsilon$) 
only, dotted black curves to
galaxy number density correlations ($nn$) 
only, and solid black curves to
results using 
all correlations
($\epsilon\epsilon$, $nn$ and $\epsilon n$). 
The grey dotted lines show results for $nn$ correlations without imposing cuts in angular frequency. The constant grey line marks the FoM computed for the pure lensing, i.e. GG, signal,
assuming intrinsic alignments do not exist. 
In addition we show the resulting figures of merit when using our most flexible parametrisation with $N_K=N_Z=7$ as filled symbols. Circles correspond to $\epsilon\epsilon$, triangles to $nn$, and diamonds to all correlations.
}
\label{fig:plot_nuispar}
\end{figure*}

In Fig.$\,$\ref{fig:plot_nuispar} both figures of merit are plotted as a function of $N_K$ and $N_Z$, respectively. If we restrict our analysis to $\epsilon \epsilon$ correlations only, our setup is similar to the most realistic setup considered in \citet{bridle07}.
We reproduce their result 
that the ${\rm FoM}_{\rm DETF}$ drops
as a function of $N_K$, dropping most sharply at small $N_K$ and then levelling off. It falls significantly below 
the reference value of the ${\rm FoM}_{\rm DETF}$ computed for a pure GG signal.
However, while \citet{bridle07} find a decrease by about a factor of 2 for $N_K=10$ and $N_Z=2$ compared to the lensing only case, our ${\rm FoM}_{\rm DETF}$ decreases by more than a factor of 4. This discrepancy can be traced back to the fact that \citet{bridle07} limit the nuisance parameter gridding to the nonlinear regime in $k$-space. 
We additionally plot the figures of merit as a function of $N_Z$, finding that the $\epsilon \epsilon$ results do in fact flatten as $N_Z$ reaches high values around 10. 
We find a very similar behaviour in terms of the ${\rm FoM}_{\rm TOT}$.
The dark energy parameters suffer more than other parameters from the uncertainty of the bias terms in the redshift direction, as the ${\rm FoM}_{\rm DETF}$ deteriorates faster than the ${\rm FoM}_{\rm TOT}$ as a function of $N_Z$.

With our default settings the 
pure $nn$ correlations constrain cosmology much more weakly than ellipticity correlations.
Recall that our galaxy clustering analysis uses a much more flexible bias parametrisation than most other work. 
Using a small number of nuisance parameters we get a ${\rm FoM}_{\rm DETF}$ which is marginally above unity. 
This result is of the same order of magnitude as the findings of \citet{rassat08} who determine ${\rm FoM}_{\rm DETF}=4.2$ for a spectroscopic, space-based survey in the spherical harmonics approach. Contrary to us, \citet{rassat08} use a less flexible bias parametrisation and include BAO features, but neglect magnification effects in their study. 
The decrease in both figures of merit for a larger number of nuisance parameters is more pronounced than for $\epsilon \epsilon$ correlations, in particular as a function of $N_Z$.

We have also
shown 
the results obtained without
the additional 
angular frequency cuts
in 
Fig.$\,$\ref{fig:plot_nuispar}
(i.e. the maximum angular frequency is $\ell_{\rm max}=3000$ for all angular power spectra). 
Because 
the full galaxy clustering signal is strong and a comparatively direct probe of the dark matter power spectrum 
if the galaxy bias is well known, 
the values of the figures of merit are much higher in this case and even surpass the lensing-only level if few nuisance parameters are used.
This corresponds to the case where we have a reasonably good understanding of galaxy biasing over a large range of scales, into the deeply non-linear regime. 
For larger $N_K$ and $N_Z$ the decrease in FoM is considerable, but weaker than with $\ell$-cuts. 

In the limit of a large number of nuisance parameters we expect that 
the galaxy bias is parametrised in a sufficiently flexible way, such that the curves with and without $\ell$-cuts should coincide or at least be of the same order of magnitude. Since this is not the case, and since there is no obvious sign of a lower boundary that the figures of merit are approaching, we hypothesise that this limit is only achieved for a very large, numerically and computationally prohibitive, number of nuisance parameters. Thus, the use of
the additional angular frequency cuts in our default analysis is the most practical way to take into account our lack of understanding of galaxy biasing on small scales. 

The simultaneous 
use of all available observables significantly boosts the parameter constraints, due to the addition of extra information from the survey, and due to breaking degeneracies between cosmological and nuisance parameters. The ${\rm FoM}_{\rm DETF}$ is up to a factor of about 50 higher than the lensing-only value. Both figures of merit decrease for larger $N_K$ and $N_Z$, attaining mostly shallow slopes at $N_K,N_Z \approx 10$. Considering the most flexible model in each panel of Fig.$\,$\ref{fig:plot_nuispar}, the ${\rm FoM}_{\rm TOT}$ remains above, but close to the value for a pure GG signal while the ${\rm FoM}_{\rm DETF}$ degrades
slightly below 
the lensing-only value. In contrast to the setup with $\epsilon \epsilon$ or $nn$ correlations only, the decrease is less pronounced with $N_Z$ than $N_K$ here, i.e. the cross-calibration between all correlations can partially compensate the loss in redshift information that affected the dark energy parameters in the former cases.

\begin{table*}[t]
\caption{Marginalised parameter errors and the two Figures of Merit for a survey with ten photometric redshift bins used for tomography $N_{\rm zbin}=10$ and a photometric redshift uncertainty parameter $\sigma_{\rm ph}=0.05$.}
\centering
\begin{tabular}[t]{ccccccccc}
\hline\hline
par. & \multicolumn{4}{c}{Euclid-like survey} & \multicolumn{4}{c}{DES-like survey}\\
 & $\epsilon \epsilon$ & $nn$ & all & lensing & $\epsilon \epsilon$ & $nn$ & all & lensing\\
\hline
$\Omega_{\rm m}$ & 0.0228 & 0.1530 & 0.0079 & 0.0044 & 0.0636 & 0.3620 & 0.0233 & 0.0120 \\
$\sigma_8$      & 0.0295 & 0.1800 & 0.0110 & 0.0065 & 0.0827 & 0.5941 & 0.0335 & 0.0180 \\  	
$h_{100}$        & 0.2072 & 0.4273 & 0.0647 & 0.1099 & 0.5447 & 1.1092 & 0.1918 & 0.2837 \\  	
$n_{\rm s}$      & 0.0721 & 0.2218 & 0.0312 & 0.0250 & 0.2117 & 1.0130 & 0.0973 & 0.0661 \\
$\Omega_{\rm b}$ & 0.0224 & 0.0329 & 0.0063 & 0.0133 & 0.0557 & 0.0771 & 0.0166 & 0.0341 \\
$w_0$           & 0.3242 & 1.6404 & 0.0939 & 0.0579 & 0.9064 & 4.3967 & 0.2895 & 0.1631 \\  	
$w_a$           & 1.1995 & 5.6187 & 0.3051 & 0.2014 & 3.4754 & 18.2522 & 0.9865 & 0.5728 \\  	
\hline
${\rm FoM}_{\rm TOT}$ & 48.30 & 24.80 & 67.13 & 66.12 & 33.64 & 7.70 & 51.87 & 52.29 \\
${\rm FoM}_{\rm DETF}$ & 9.12 & 0.28 & 102.92 & 182.96 & 1.08 & 0.04 & 10.96 & 24.92 \\  
\hline
\end{tabular}
\tablefoot{We have used the configuration with the maximum set of nuisance parameters that we consider, i.e. $N_K=N_Z=7$. The cuts in angular frequency for number density correlations have been applied to restrict the analysis to the linear regime. Shown are the results using shear-shear correlations $\epsilon \epsilon$ only, observed galaxy clustering correlations $nn$ only, all correlations, and the lensing-only signal (assuming no intrinsic alignments). The left-hand part of the table uses the standard Euclid-like survey parameters, i.e. $\sigma_\epsilon=0.35$, $n=35\,{\rm arcmin}^{-2}$, and the survey area $A=20000\,{\rm deg}^2$. The right-hand part of the table uses parameters for a DES-like survey, i.e. $\sigma_\epsilon=0.23$, $n=10\,{\rm arcmin}^{-2}$, and the survey area $A=5000\,{\rm deg}^2$.}
\label{tab:results}
\end{table*}

The most 
flexible configuration we consider
has 
$N_K=N_Z=7$ which corresponds to 200 nuisance parameters within the four bias terms. This limit is not inherent to our method, but is merely set
for computational practicality. 
Both the intrinsic alignments and the galaxy bias have a physical origin and therefore
are expected to 
produce smooth signals, which should not oscillate strongly or feature sharp peaks. Hence
on the default angular scales used 
this 
model with 100 free parameters for
each of 
intrinsic alignments and galaxy bias
should yield a fairly general representation of the signals if one can rely on coarse prior information on the fiducial base model.

\begin{figure*}[ht]
\centering
\includegraphics[scale=.87,angle=270]{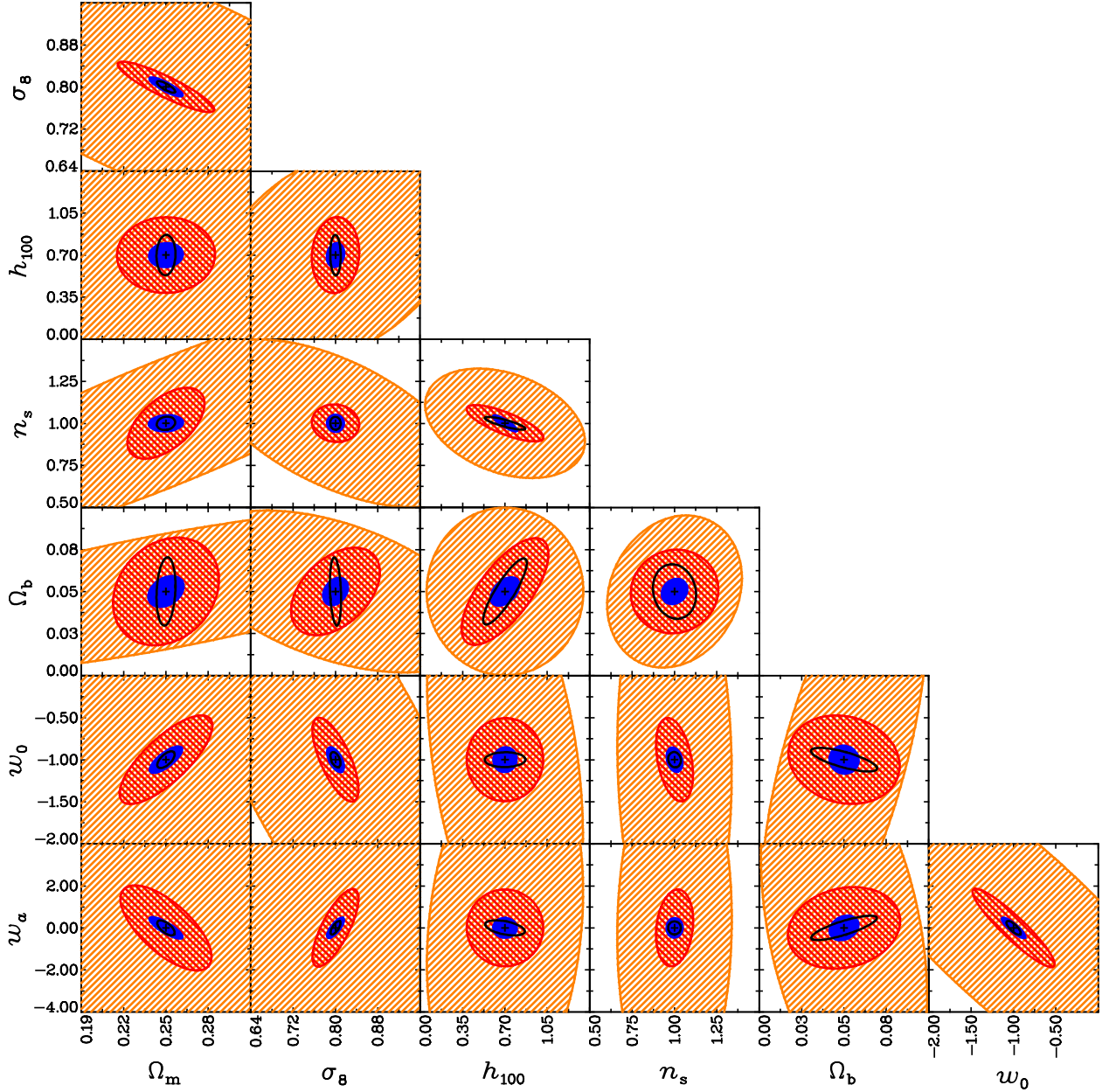}
\caption{
$1\,\sigma$-contours for all pairs of cosmological parameters considered,
marginalised over all other parameters. 
We have used
a photometric redshift uncertainty parameter value 
$\sigma_{\rm ph}=0.05$,
ten photometric redshift bins for tomography 
$N_{\rm zbin}=10$,
and the most flexible intrinsic alignment and bias model considered in this paper, with over two hundred free parameters ($N_K=7$, and $N_Z=7$). 
Orange 
(light hatched) 
confidence regions result from using galaxy number density correlations ($nn$) (excluding the
non-linear regime) only, red 
(dark hatched) 
regions
use 
ellipticity correlations ($\epsilon \epsilon$)
alone, 
and blue 
(filled) 
regions correspond to using all available information including density-ellipticity cross-correlations.
For reference, the contours obtained from a pure lensing signal are shown as black lines. 
Flat 
priors on cosmological parameters have been applied.
}
\label{fig:contours}
\end{figure*}

The figures of merit and the individual parameter errors for the most flexible model are given in Table \ref{tab:results}, for the pure GG signal, $\epsilon \epsilon$, $nn$, and all correlations. Compared to lensing alone, the ${\rm FoM}_{\rm DETF}$ decreases by about
a factor of 
20 for $\epsilon \epsilon$ correlations, which means that cosmic shear is severely affected if one assumes very little prior knowledge about intrinsic alignments. There are hardly any dark energy constraints for $nn$ correlations only, given this freedom in the galaxy bias. However, using all available correlation simultaneously, we can recover
just over half 
of the pure
cosmic shear 
DETF 
figure of merit. We find that the total error volume of the cosmological parameters, i.e. the ${\rm FoM}_{\rm TOT}$, is the same
using all the correlation information 
as for the lensing-only case. Looking at the marginalised parameter errors, $w_0$ and $w_a$ suffer particularly strongly in the $\epsilon \epsilon$ only and $nn$ only case. Less information about the dark energy parameters is lost if all correlations are used, and
the constraints on $h_{100}$ and $\Omega_{\rm b}$ improve over the lensing-only case.
In Fig.$\,$\ref{fig:contours} we show in addition the marginalised $1\,\sigma$-error ellipses of all possible pairs of cosmological parameters for the same setup.
The degeneracy directions are fairly similar for the usual lensing only case and the results with all correlations presented in this paper. 

We repeat this analysis for a
nearer term Stage-III-like 
survey
such as the Dark Energy Survey (DES). 
To this end, we use a survey size of $A=5000\,{\rm deg}^2$, a median redshift of the overall redshift distribution of $z_{\rm med}=0.8$, a total number density of galaxies of $n=10\,{\rm arcmin}^{-2}$, and a total ellipticity dispersion of $\sigma_\epsilon=0.23$, keeping all other parameters at their default values.
We refer to this survey as DES-like for the rest of the paper. 

Note that different groups use different conventions for quoting the number density of galaxies and the ellipticity dispersion. The Euclid team define $\sigma_\epsilon=0.35$, which is motivated by considering it to be the typical uncertainty on the shear taking into account both shape noise and measurement errors. The effective number density has then been estimated relative to this value and found to be between 30 and 40$\,{\rm arcmin}^{-2}$ \citep{laureijs09}. Because of the choice of $\sigma_\epsilon$, this number density value is approximately equal to the actual number of usable galaxies in the image. The DES team start by defining $\sigma_\epsilon=0.23$, motivated by considering the shape noise alone. The effective number density is then found to be $n=10\,{\rm arcmin}^{-2}$ \citep{annis05}. Due to this choice of $\sigma_\epsilon$, the number density value is smaller than the number density of usable galaxies in the image, and represents the effective number density of galaxies with negligible measurement errors. The different definitions of $\sigma_\epsilon$ also explain the low number density quoted for DES in comparison with shallower surveys such as the CHFTLS.

In this work we choose to take the number density and ellipticity dispersion values as provided by the collaborations and do not adjust them, which would be beyond the scope of this paper. In any case the physically meaningful quantity is the ratio $\sigma_\epsilon/\sqrt{n}$ which determines the uncertainties on the power spectra, see (\ref{eq:covnoiseterms}). Thus, as long as we use consistent definitions of $n$ and $\sigma_\epsilon$, we obtain the correct shape noise contributions to the cosmic shear covariance.

Our findings are also shown in Table \ref{tab:results}. For the DES-like survey the ${\rm FoM}_{\rm TOT}$ for all correlations is again about the same as in the lensing-only case,
and the ratios of ${\rm FoM}_{\rm DETF}$ values are 
slightly smaller than for 
the Stage-IV-like survey.
The ${\rm FoM}_{\rm DETF}$ values 
for $\epsilon \epsilon$, $nn$, and all possible correlations 
are about a factor of ten larger from the Stage-IV-like cosmic shear survey than the Stage-III-like survey, offering significant benefit beyond the minimal requirement of \citet{albrecht06}. 

\subsection{Dependence on characteristics of the redshift distribution}
\label{sec:resultsredshift}

It is well known that dividing the galaxy sample into several redshift
photo-z 
bins 
greatly improves constraints
from 
cosmic shear, but due to the broad lensing kernel (\ref{eq:weightlensing}) there is little benefit in having more distributions than three to five \citep[e.g.][]{hu99,simon04,ma05}. This result does not hold true anymore if one aims at controlling the intrinsic alignment contamination in the cosmic shear signal, mainly manifest via its characteristic redshift dependence. Using both marginalisation \citep{bridle07} and parameter-free approaches \citep{joachimi08b}, one finds that the figures of merit only start to stabilise when using ten redshift distributions or more.

\begin{figure}[t]
\centering
\includegraphics[scale=.6]{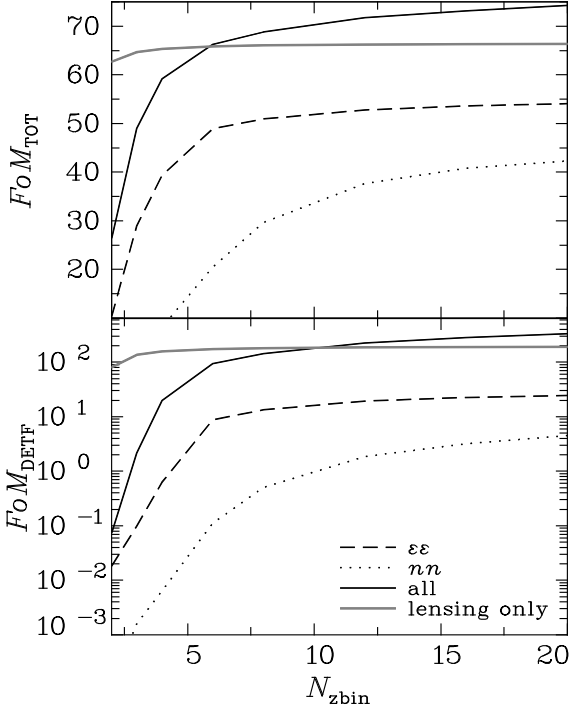}
\caption{\textit{Upper panel}: The
Figure of Merit for all cosmological parameter space 
${\rm FoM}_{\rm TOT}$ as a function of the number of photo-z bins
used for tomography 
$N_{\rm zbin}$, shown for $\epsilon \epsilon$ (dashed line), $nn$ (dotted line), and all (solid) correlations. The grey line corresponds to results for lensing only (GG). Throughout, $N_K=N_Z=5$ nuisance parameters for the bias terms are used. These results are obtained for the standard set of parameters and $\sigma_{\rm ph}=0.05$. \textit{Lower panel}: Same as above, but in terms of
the dark energy figure of merit 
${\rm FoM}_{\rm DETF}$.}
\label{fig:plot_nz}
\end{figure}

In Fig.$\,$\ref{fig:plot_nz} our figures of merit are plotted as a function of the number of photo-z bins, using $N_K=N_Z=5$ nuisance parameters per bias term. In agreement with the findings mentioned above both the ${\rm FoM}_{\rm TOT}$ and the ${\rm FoM}_{\rm DETF}$ in the case of a pure cosmic shear signal become approximately constant for $N_{\rm zbin} \gtrsim 5$, and increasing $N_{\rm zbin}$ beyond three has little effect. For $\epsilon \epsilon$ correlations the gain in FoM is considerable up to $N_{\rm zbin} \sim 7$; for larger $N_{\rm zbin}$ the curves rise only slowly.

Since $nn$ correlations are more localised because of the compact kernel of the dominating galaxy clustering (gg) signal, increasing $N_{\rm zbin}$ proves beneficial for these observables up to the maximum number of photo-z bins we have considered, although the bins feature an increasing overlap of their corresponding redshift distributions as we keep $\sigma_{\rm ph}=0.05$ fixed. As one would expect, we obtain an intermediate scaling with $N_{\rm zbin}$ for the complete set of available correlations. Our fiducial choice of $N_{\rm zbin}=10$ is beyond the regime of strongly varying figures of merit at small $N_{\rm zbin}$, but the further increase in FoM is more pronounced than for a pure lensing signal or $\epsilon \epsilon$ correlations only
with ${\rm FoM}_{\rm DETF}$ rising by an additional $80\,\%$ on increasing the number of photometric redshift bins from 10 to 20. 

\begin{figure}[t]
\centering
\includegraphics[scale=.6]{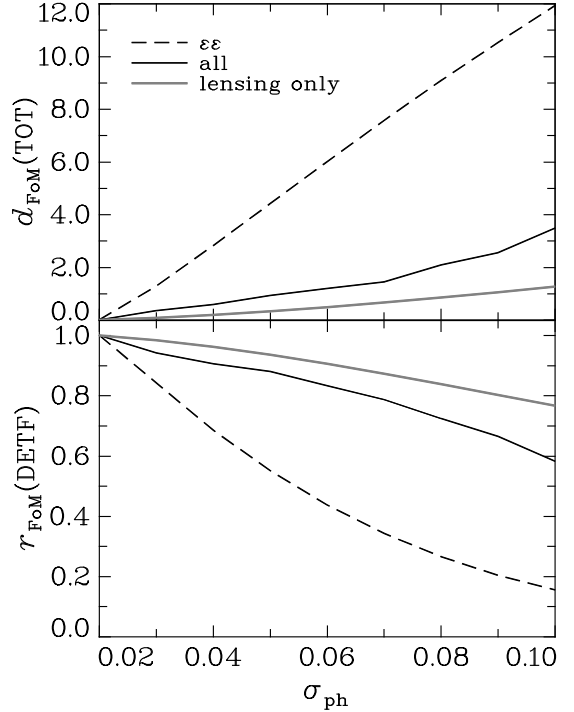}
\caption{\textit{Upper panel}: The difference $d_{\rm FoM}$, defined in (\ref{eq:fomrelative}), as a function of the photo-z dispersion $\sigma_{\rm ph}$, shown for $\epsilon \epsilon$ correlations (dashed line), all correlations (black solid line), and the lensing only signal (grey solid line). Throughout, nuisance parameters $N_K=N_Z=5$ are used. These results are obtained for the standard set of parameters and $N_{\rm zbin}=10$. \textit{Lower panel}: Same as above, but in terms of the ratio $r_{\rm FoM}$, given in (\ref{eq:fomrelative2}).} 
\label{fig:plot_sigma}
\end{figure}

Figure \ref{fig:plot_sigma} shows the figures of merit as a function of the photo-z dispersion, normalised to the value at $\sigma_{\rm ph}=0.02$.
Since the ${\rm FoM}_{\rm TOT}$ is a logarithmic quantity, we compute differences rather than ratios of the figure of merit, i.e. 
\eqa{
\label{eq:fomrelative2}
r_{\rm FoM} &=& {\rm FoM}_{\rm DETF}(\sigma_{\rm ph})/{\rm FoM}_{\rm DETF}(\sigma_{\rm ph}=0.02)\;\\ 
\label{eq:fomrelative}
d_{\rm FoM} &=& {\rm FoM}_{\rm TOT}(\sigma_{\rm ph}=0.02) - {\rm FoM}_{\rm TOT}(\sigma_{\rm ph})\\ \nn
&=& \ln \br{ \frac{ \bb{ \det \br{F^{-1}}_{\rm cosm.}}(\sigma_{\rm ph})}{\bb{ \det \br{F^{-1}}_{\rm cosm.}}(\sigma_{\rm ph}=0.02)} }\;.
}
Hence the difference $d_{\rm FoM}$ is directly related to the change in volume of the error ellipsoid spanned by the set of cosmological parameters. 
We have
returned to our default value of 
$N_{\rm zbin}=10$.

The pure lensing signal needs merely coarse redshift information to attain its full statistical power, and hence its FoM hardly suffers from the increasing spread in the redshift distributions. However, as redshift information is vital to account for intrinsic alignments, the figures of merit for the cosmic shear signal after marginalisation over intrinsic alignments decrease considerably by
more than $80\,\%$ in terms of the ${\rm FoM}_{\rm DETF}$  
on changing from $\sigma_{\rm ph}=0.02$ to $\sigma_{\rm ph}=0.1$, which 
is in line with \citet{bridle07}. For the same change in $\sigma_{\rm ph}$ the ${\rm FoM}_{\rm TOT}$ is reduced by about 12.
It is interesting to note that the dependence of the ${\rm FoM}_{\rm TOT}$ on $\sigma_{\rm ph}$ is close to linear for both lensing and $\epsilon \epsilon$ signal. 

Adding galaxy number density information largely alleviates the information loss. The degradation in FoM is only slightly stronger than for the pure lensing signals in the expected regime of high-quality photo-z information with $\sigma_{\rm ph} \lesssim 0.06$. For larger $\sigma_{\rm ph}$ the increasing overlap of the redshift distributions causes both figures of merit to
decrease further. 
We have to add the caveat that in this investigation we consider only a single parameter
that accounts for the uncertainty in the shape of the redshift distributions,
$\sigma_{\rm ph}$. 

It remains to be seen how sensitive these conclusions are to greater levels of uncertainty in the photometric redshift distribution calibration. For example \citet{zhang09} consider the use of nn and $\epsilon$n terms for self-calibrating these uncertainties but do not include intrinsic alignments in their main calculations. 
If we assumed e.g. uncertain median redshifts of each individual distribution, which is beyond the scope of this paper and under investigation elsewhere, we would have obtained significantly lower figures of merit; see 
\citet{kitching08c} and also 
\citet{bridle07} who consider this for the case of $\epsilon \epsilon$ correlations only. The curve for all correlations in Fig.$\,$\ref{fig:plot_sigma}
may 
hence approach the one for $\epsilon \epsilon$ correlations, especially for $\sigma_{\rm ph} \lesssim 0.06$. Consequently, redshift distributions with a small spread
may turn out to be even more 
desirable when taking galaxy number density information into account.

\subsection{Dependence on nuisance parameter priors}
\label{sec:resultspriors}

\begin{figure}[t]
\centering
\includegraphics[scale=.6]{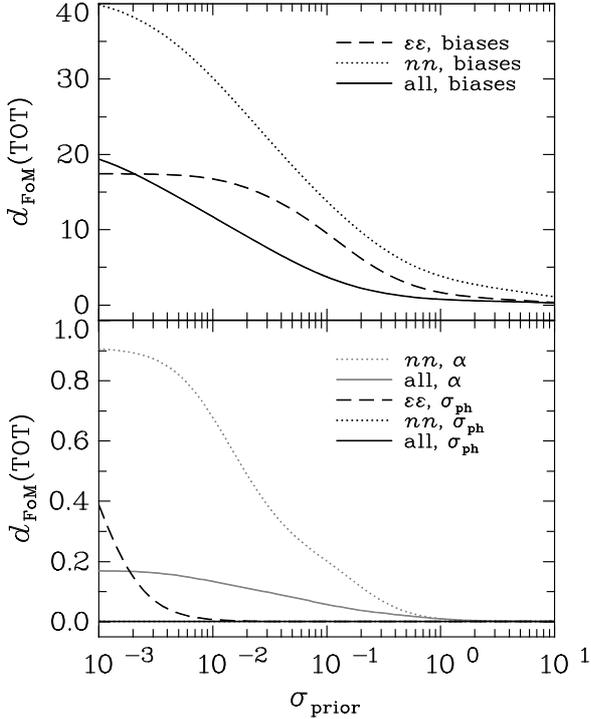}
\caption{
The difference $d_{\rm FOM}$ between the ${\rm FoM}_{\rm TOT}$ for a given prior value $\sigma_{\rm prior}$ and the fiducial ${\rm FoM}_{\rm TOT}$, obtained for the least stringent, default values of the different priors we apply. 
\textit{Upper panel}: Effect of tightening the priors on the nuisance parameters in the bias terms for $\epsilon \epsilon$ (dashed line), $nn$ (dotted line), and all (solid) correlations. \textit{Lower panel}: Effect of tightening the priors on the slopes of the luminosity function $\alpha^{(i)}$, and for the prior on the uncertainty of $\sigma_{\rm ph}$. Grey curves correspond to priors on $\alpha^{(i)}$, black curves to those on $\sigma_{\rm ph}$. 
Except for the $\epsilon \epsilon$ case the latter remain very close to zero. 
As above, $\epsilon \epsilon$ correlations are shown as dashed lines, $nn$ correlations as dotted lines, and all correlations as solid lines. Note that $\epsilon \epsilon$ correlations do not depend on the $\alpha^{(i)}$. For all curves, the remaining priors are each set to their default values of 50 for the bias term parameters, and 10 for priors on both $\alpha^{(i)}$ and $\sigma_{\rm ph}$.}
\label{fig:plot_priors}
\end{figure}

On all our nuisance parameters we can expect to have prior information to a certain extent, at least by the time large space-based cosmological surveys will be undertaken. Here we investigate the dependence of the resulting ${\rm FoM}_{\rm TOT}$ on tightening the priors on the different sets of nuisance parameters, employing our most general configuration with $N_K=N_Z=7$. We show 
$d_{\rm FoM} = {\rm FoM}_{\rm TOT}(\sigma_{\rm prior}) - {\rm FoM}_{\rm TOT}(\sigma_{\rm prior}^{\rm max})$, i.e. 
the ${\rm FoM}_{\rm TOT}$, referred to its value for the widest priors we apply as default, as a function of the
Gaussian prior width $\sigma_{\rm prior}$ 
in Fig.$\,$\ref{fig:plot_priors}.

For reasons of numerical stability, we have imposed wide Gaussian priors on all nuisance parameters, i.e. 10 on both the $\alpha^{(i)}$ and $\sigma_{\rm ph}$, and 50 on the parameters within the
intrinsic alignment and galaxy 
bias terms. As the curves clearly indicate, these priors are non-informative
since all of the curves have flattened off by this value. 
The strongest effect is achieved by tightening the priors on the bias terms (note the different scaling of the two panels). Prior knowledge on the galaxy bias tremendously improves constraints by $nn$ correlations, raising the ${\rm FoM}_{\rm TOT}$ close to the values of the lensing-only case if the uncertainty of the bias term nuisance parameters is of the order $10^{-3}$. Since in a situation with such excellent prior knowledge about galaxy bias, the cuts in angular frequency could be much less stringent, the increase in FoM would be even more pronounced.

The ${\rm FoM}_{\rm TOT}$ for $\epsilon \epsilon$ correlations changes from its value at the default configuration to the lensing-only value\footnote{In fact we expect the ${\rm FoM}_{\rm TOT}$ for $\sigma_{\rm prior} \rightarrow 0$ to be slightly above the lensing-only case since the intrinsic alignment signals also constrain cosmology if their form is perfectly known, see also \citet{bridle07}.} (compare to Table \ref{tab:results}) in the limited range between $\sigma_{\rm prior} \sim 1$ and $\sigma_{\rm prior} \sim 10^{-2}$. Thus coarse prior information about intrinsic alignments has little effect on the FoM, but, compared to the galaxy bias, a model with comparatively moderate precision can already bring the ${\rm FoM}_{\rm TOT}$ back close to its optimum. If all available correlations are used, the internal calibration works well to constrain the nuisance parameters, so that the priors only mildly improve the ${\rm FoM}_{\rm TOT}$. However, bear in mind that in spite of our flexible parametrisation, the choice of a fiducial base model for both intrinsic alignments and galaxy bias influences the results. Hence the base models should be as realistic as possible. It is questionable whether our knowledge about the bias terms will ever suffice to impose priors on the bias term nuisance parameters beyond the approximately correct form of the base models.

As already discussed in the foregoing section, the global parameter $\sigma_{\rm ph}$ is excellently constrained by the $nn$ correlations. Therefore even tight priors do not have any effect on the ${\rm FoM}_{\rm TOT}$ if $nn$ correlations form part of the data vector. Priors below $\sigma_{\rm prior}=10^{-2}$ increase the information content in $\epsilon \epsilon$ correlations marginally. As far as the uncertainty in the redshift distributions is concerned, our investigation is still to be regarded as idealistic. We defer the joint analysis of shape and number density correlations in presence of unknown parameters in each individual redshift distribution to future work.

The lack of knowledge in the slopes of the luminosity function $\alpha^{(i)}$ entering the lensing magnification signal is negligible in comparison with the effect of the galaxy bias nuisance parameters. If all correlations are considered, the $\alpha^{(i)}$ are well constrained since the improvement in FoM due to priors is below the $1\,\%$ level.

\subsection{Information content in the individual signals}
\label{sec:resultssubsignals}

The question arises
of 
which signals
contributing 
to the observable power spectra contain most of the information about cosmology or suffer
most strongly 
from the uncertainty in intrinsic alignments and galaxy bias. \citet{bridle07} have studied the effect of the two intrinsic alignment terms on $\epsilon \epsilon$ correlations with a parametrisation very similar to ours. In the following we consider $nn$ correlations by repeating the Fisher matrix analysis twice: once assuming that there is no lensing magnification signal, i.e.
only 
gg
contributes to nn; 
and once assuming that there is no intrinsic galaxy clustering, i.e.
only 
mm
contributes to nn. 

\begin{figure}[t]
\centering
\includegraphics[scale=.6,angle=270]{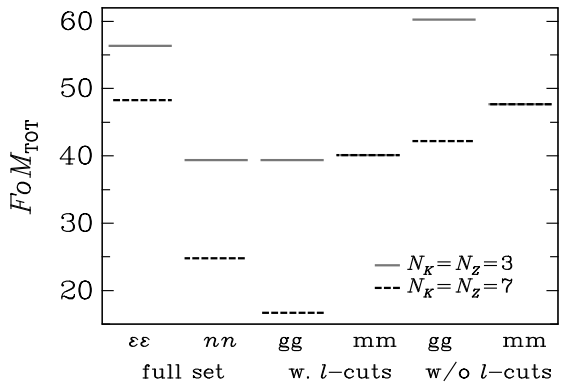}
\caption{The ${\rm FoM}_{\rm TOT}$ for different subsets of correlations that form the $nn$ signal. Throughout, we used $N_{\rm zbin}=10$ and $\sigma_{\rm ph}=0.05$. The marginalisation has been performed for $N_K=N_Z=7$ nuisance parameters (black dashed bars) or $N_K=N_Z=3$ nuisance parameters (grey solid bars). The two leftmost columns show the full set of $\epsilon \epsilon$ and $nn$ correlations for reference. The two centre columns stand for gg and mm correlations only, employing the cuts in angular frequency. The rightmost columns show again gg and mm, but here the full information up to $\ell_{\rm max}=3000$ is considered. Note that mm correlations depend neither on intrinsic alignments nor on galaxy bias and hence are independent of $N_K$ and $N_Z$.}
\label{fig:plot_subsets_nn}
\end{figure}

For this calculation we consider constraints from galaxy clustering information alone ($nn$), and assume that galaxy shape information ($\epsilon \epsilon$ and $n\epsilon$) is not used. 
We show the resulting ${\rm FoM}_{\rm TOT}$ for the most flexible model ($N_K=N_Z=7$) and a model with
somewhat less flexibility ($N_K=N_Z=3$) 
in Fig.$\,$\ref{fig:plot_subsets_nn}.
On comparing the second, third and fourth columns from the left we see that 
the model with less freedom
(solid grey lines) 
has approximately the same 
${\rm FoM}_{\rm TOT}$
for the complete $nn$ signal
as when the $nn$ signal is made up of either one of the 
pure gg and mm terms
alone. We now examine how this conclusion is changed when increased flexibility is allowed in the galaxy bias model (dashed lines). 
Lensing magnification does not depend on any of our bias terms, so that its ${\rm FoM}_{\rm TOT}$ value is the same for the flexible model
(the dashed and solid lines are on top of each other in the fourth column from the left). 
In contrast,
constraints on cosmology from the intrinsic galaxy clustering information alone (gg) weaken significantly as greater flexibility is included in the galaxy bias model. Specifically, 
the
constraint from the 
gg signal drops by more than 20 in the logarithmic ${\rm FoM}_{\rm TOT}$
when increasing the number of nuisance parameters per bias term by a factor of five
(compare the solid and dashed lines in the third column from the left). 

The $nn$ correlations, containing both clustering and magnification, unsurprisingly have a ${\rm FoM}_{\rm TOT}$ in between
(second column from the left). 
Hence, if the galaxy bias is well known, the galaxy clustering signal dominates the
information from the 
$nn$ correlations, but
for very flexible bias
models
lensing magnification
does allow more information on cosmology to be extracted, as compared to what might be expected if magnification did not take place (compare the dashed lines in the second and third columns).

The two right hand columns of Fig.$\,$\ref{fig:plot_subsets_nn} show 
the ${\rm FoM}_{\rm TOT}$ obtained
when the full range of angular scales is used in the power spectra, thus including information from non-linear scales. 
As already seen in Fig.$\,$\ref{fig:plot_nuispar}, the clustering constraints improve strongly when adding the signal from the non-linear regime.
The magnification contribution remains trustworthy far into the non-linear regime, but is of course also affected by the $\ell$-cuts.
When using the full range of angular scales the ${\rm FoM}_{\rm TOT}$ of the mm signal is increased by about 7. 
The cosmological constraints from intrinsic galaxy clustering (gg) alone are now much tighter than from the magnification effect alone (mm), when the smaller number of bias parameters are used (compare solid lines in the two right hand columns). However, the cosmological constraints from the maximally flexible model are still tighter from magnification alone than from galaxy clustering (dashed lines in the right hand two columns). 

Note that the
constraints from the 
isolated mm and gg signals are hypothetical -- the full $nn$ correlations
including both terms 
are the only true observables. It is
only 
possible to separate the contributions in an approximate fashion when making use of their characteristic scaling with redshift, see Fig.$\,$\ref{fig:plotps}.
But all of this available information is already included in our $nn$ results when a large enough number of tomographic redshift bins are used. 

We now 
assess the
cosmological information available in 
the different subsets of observables when performing the Fisher matrix analysis for all correlations simultaneously. To this end, we split up the summation of the Fisher matrix (\ref{eq:fishermatrix}) into three parts, corresponding to $\epsilon \epsilon$, $nn$, and $n \epsilon$ correlations.
We also consider pairs of observables e.g. $\epsilon \epsilon$, and $nn$. 
The Fisher matrices of each part
are then inverted separately to yield individual parameter errors and figures of merit. Thereby we split the total information into subsets of the data vector, which could in principle be observed independently.

However, the figures of merit we compute do not correspond to those which one would obtain for an independent analysis of $\epsilon \epsilon$, $nn$, or $n \epsilon$ correlations because
we extract the relevant rows and columns of the inverse of the full covariance matrix (\ref{eq:covtotal}) to insert into the Fisher matrix calculation (\ref{eq:fishermatrix}). 
Due to the inversion, the covariance terms of the different subsets mix, thus accounting for the cross-correlations between the observables in the different subsets.
This 
is desirable for our purposes
because these terms add together in the full calculation. 
Consequently, it is possible that the FoM of a subset is larger than the one obtained for the complete data vector if there are anti-correlations with observables of other subsets which are not taken into account due to the splitting. Formally speaking, this means that due to anti-correlations of different observables, off-diagonal terms of the Fisher matrix can have negative entries, which produce negative terms in the sum in (\ref{eq:fishermatrix}). We indeed observe this behaviour for one of the subsets.

\begin{figure}[t]
\centering
\includegraphics[scale=.6]{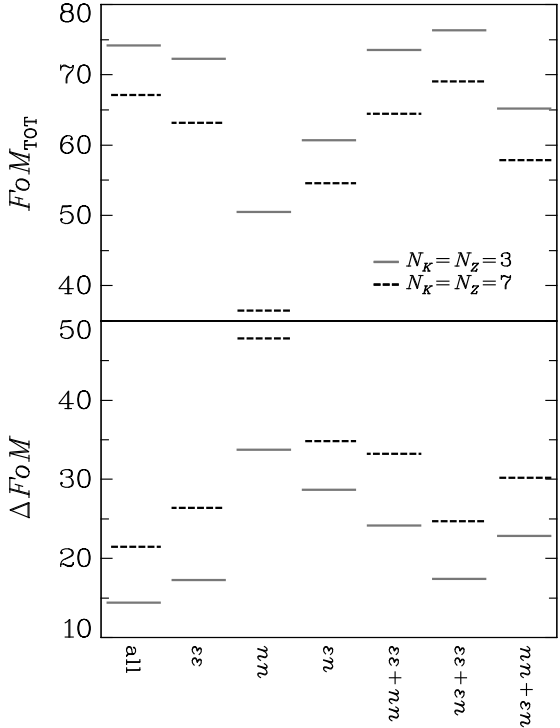}
\caption{\textit{Upper panel:} The ${\rm FoM}_{\rm TOT}$ for different subsets of correlations that are observable from a galaxy survey with galaxy shape and number density measurements. The Fisher matrix has been computed for the total data set containing all correlations, i.e. these results represent the information content of the subsets as a part of the total signal. Black dashed bars correspond to a marginalisation over $N_K=N_Z=7$ nuisance parameters, grey solid bars to $N_K=N_Z=3$ nuisance parameters. The labels on the abscissa indicate the different combinations of correlations used, where $\epsilon n$ stands for the cross-correlations between number density and ellipticity. \textit{Lower panel:} Difference $\Delta$ FoM, see (\ref{eq:deltafom}), for the same subsets as above. This difference can be understood as a measure of the depletion of information in the subsets due to the marginalisation over nuisance parameters.}
\label{fig:plot_subsets_all}
\end{figure}

The ${\rm FoM}_{\rm TOT}$ for the subsets of $\epsilon \epsilon$, $nn$, and $n \epsilon$ correlations, as well as all possible combinations thereof, are given in
the upper panel of 
Fig.$\,$\ref{fig:plot_subsets_all}, again for the two parametrisations $N_K=N_Z=\bc{3;7}$. As part of the full set of observables, the $\epsilon \epsilon$ correlations, governed by the cosmic shear signal, yield the highest ${\rm FoM}_{\rm TOT}$ and contribute most to the cosmological information in the full set
(second column). 
They are followed by cross-correlations between number density and ellipticity, which have a moderate ${\rm FoM}_{\rm TOT}$
(fourth column). 
This occurs 
in spite of the
smaller range in 
angular frequency
used for this observable. 
In addition, the least amount of information is lost 
when switching to the more flexible bias
model. The subset of $nn$ correlations has by far the lowest ${\rm FoM}_{\rm TOT}$, which becomes even more pronounced when the bias model has more nuisance parameters. This hierarchy in FoM is also mirrored in the results for the different combinations of two subsets.

To get a more explicit measure of the effect of the marginalisation over the galaxy bias and intrinsic alignments nuisance parameters, we compute the quantity
\eqa{
\label{eq:deltafom}
\Delta {\rm FoM} &\equiv& \ln \bb{\det F_{\rm cosm.}} - {\rm FoM}_{\rm TOT}\;\\ \nn
&=& - \ln \bb{\det \br{ F_{\rm cosm.} }^{-1}} + \ln \bb{\det \br{F^{-1}}_{\rm cosm.}}\;
}
for each of the subsets. The first term in (\ref{eq:deltafom}) is of a similar form as the ${\rm FoM}_{\rm TOT}$, but
the cosmological elements are extracted before the inverse or determinant is taken 
i.e. it does not include the marginalisation over nuisance parameters. 
Hence, $\Delta {\rm FoM}$ quantifies the depletion in FoM due to the marginalisation. As is evident from the lower panel in Fig.$\,$\ref{fig:plot_subsets_all}, the susceptibility of the subsets to the nuisance parameters is closely related to their contribution to the ${\rm FoM}_{\rm TOT}$ of the full set. Accordingly, $\Delta {\rm FoM}$ is largest for the $nn$ subset while the ellipticity-number density cross-correlations have substantially smaller $\Delta {\rm FoM}$ although they contain all four bias terms. The effect of intrinsic alignments on the $\epsilon \epsilon$ subset is relatively small
when compared to the effect of galaxy bias on the $nn$ correlations. 

Since the
$\epsilon \epsilon$ 
correlations only depend on
intrinsic alignments 
($b_{\rm I}$ and $r_{\rm I}$) 
whereas $nn$ correlations only feature
galaxy bias 
($b_{\rm g}$ and $b_{\rm g}$), 
combining them helps little in reducing $\Delta {\rm FoM}$, see the column labelled \lq$\epsilon \epsilon+nn$\rq.
Both lines in this column remain higher than for the other combinations \lq$\epsilon \epsilon+\epsilon n$\rq and \lq$nn + \epsilon n$\rq. 
The ellipticity-number density cross-correlations instead have great potential in breaking degeneracies between cosmological and nuisance parameters. In particular, adding their subsets to the $\epsilon \epsilon$ correlations further decreases $\Delta {\rm FoM}$ below the $\epsilon \epsilon$-only value for the most flexible bias term model. This synergy is presumably related to the internal calibration between the IG and gI signals, as investigated by \citet{zhang08}. Finally, the full set of shape and number density observables clearly calibrates the nuisance parameters best as it produces the smallest $\Delta {\rm FoM}$ for both the most flexible and the more rigid bias term models.

\section{Conclusions}
\label{sec:conclusions}

In this work we studied the joint analysis of galaxy number density and shape correlations to constrain cosmological parameters in presence of contaminations by the intrinsic alignment of galaxies and the galaxy bias. We considered the signals due to gravitational shear, intrinsic shear, intrinsic galaxy clustering, and lensing magnification, explicitly computing all possible two-point correlations thereof. We introduced a two-dimensional grid parametrisation to account for the unknown
scale 
and redshift dependence of both intrinsic alignments and galaxy bias. Further nuisance parameters were used to describe the uncertainty in the width of the photo-z bin-wise redshift distributions and in the slope of the galaxy luminosity functions within each bin.

Our Fisher matrix analysis demonstrates that the simultaneous use of ellipticity correlations, number-density correlations, and in particular ellipticity-number-density cross-correlations allows for a substantial amount of internal calibration of the bias terms. With flexible models that contain in total more than 200 nuisance parameters we can recover the volume of the error ellipsoid in parameter space when compared to assuming a pure gravitational lensing signal
and using just the ellipticity correlation information. 
The dark energy parameters $w_0$ and $w_a$
suffer more than other parameters on 
marginalisation over nuisance parameters, so that $56\,\%$ of the ${\rm FoM}_{\rm DETF}$ are retained for a Euclid-like survey in this most flexible setup
considered. 
The ${\rm FoM}_{\rm DETF}$ for the combined set of shape and number density correlations is close to the pure lensing ${\rm FoM}_{\rm DETF}$ if we choose a model which uses about 100 nuisance parameters to describe intrinsic alignments and galaxy bias. Our approach also proves beneficial for upcoming ground-based surveys 
with DES-like survey characteristics. 
The benefit is greatest for the more ambitious survey. 
In addition, we assumed the slopes of the galaxy luminosity functions in each photo-z bin to be unknown nuisance parameters and found that they are well calibrated internally from the data.

The information which we added on top of the standard cosmic shear analysis comes without any extra cost since galaxy number density measurements are directly available from imaging data. Given our encouraging findings, we hence suggest that the joint consideration of galaxy shape and number density information could become the standard technique whenever intrinsic alignments are suspected to make a significant contribution to the cosmic shear signal.

When interpreting our results, one has to keep in mind that our grid parametrisation of the bias terms has limited flexibility. 
On increasing the number of nuisance parameters in the grids, the curves for the figures of merit flatten off. However we are at present not able to say whether they approach a lower limit or continue to decrease for a large number of grid nodes. This issue, which is of considerable theoretical and practical interest, is currently under investigation, as well as a comparison with the performance of removal techniques such as nulling \citep{joachimi08b,joachimi09}. 

Due to the finite number of grid points the resulting angular power spectra will not span perfectly the range of possible physical models. 
Moreover, we found that of order 10 nuisance parameters in $k$ per grid node in redshift do not suffice to represent all relevant functional forms of the galaxy bias. Therefore we removed all observables from the parameter estimation which have a significant contribution from the galaxy clustering signal 
at large $k$ 
where the signal is strongest and the bias least known. However, since in the near future we expect to have at least coarse, but reliable knowledge about the functional forms of the galaxy bias on linear scales and the intrinsic alignment signals, the models used in our approach should still yield realistic results.

While the log-linear grid parametrisation is fairly general and intuitive, it may not be the most efficient way to represent freedom in intrinsic alignments and the galaxy bias. As all bias terms originate from physical processes, they are smooth and do not feature strong oscillations or isolated peaks. Thus, we presume that the bias terms can efficiently be parametrised in terms of complete sets of smooth functions such as the Fourier, Chebyshev, or Legendre series. Truncating the higher orders of these series will only limit the model to represent highly oscillatory or small-scale features. These parametrisations might therefore be more efficient in comprising the full set of realistic bias terms \citep[see also][]{kitching08} for a given number of nuisance parameters, this number in turn being dictated by the available computational power. 
The latter will play a critical role when the approach suggested here is performed in a full likelihood analysis with several hundreds of nuisance parameters. 
We add that it might be necessary to furthermore consider the bias terms individually for different galaxy types
and luminosities 
because 
it is known 
that both the intrinsic alignments and the galaxy bias vary considerably with galaxy type
and luminosity 
\citep[for recent examples see][]{mandelbaum06,hirata07,mccracken07, swanson08, simon08, cresswell09, wang07}. 

Redshift information is crucial
for discriminating 
the different signals that contribute to the observables. We investigated the dependence of the parameter constraints
from 
the joint set of correlations on characteristics of the redshift distributions. We confirm the observation by foregoing works that the number of photo-z bins needed to retrieve the bulk of information about cosmology increases by at least a factor of two when
using only 
ellipticity correlations
and marginalising 
over the intrinsic alignment signals. Using the complete set of correlations, the figures of merit do not level off
so quickly, 
but continue to increase
as the number of tomographic bins is increased. 
This can be explained by the narrower kernel in the redshift integrations that link the projected and the three-dimensional power spectra involving galaxy number-density signals.

Moreover the figures of merit decrease substantially
more slowly 
as a function of the photometric redshift dispersion $\sigma_{\rm ph}$ when adding galaxy number density information to the data. However, in our marginalisation we only used a single parameter accounting for the uncertainty in the spread of all bin-wise redshift distributions. This way the redshift cross-terms of the number-density correlations efficiently calibrate the shape of the redshift distributions. Therefore this result
alone cannot 
be interpreted as a potential relaxation of the requirements for photometric redshift accuracy in cosmological surveys featuring cosmic shear. Conversely, the continuing increase in the figure of merit as a function of $N_{\rm zbin}$ makes an even larger number of well-separated and compact redshift distributions desirable in the case of the joint data set.

The recent work by~\citet{zhang09} uses density-ellipticity correlations to self-calibrate photometric redshift parameters but a full approach may need to simultaneously deal with intrinsic alignments and photometric redshift properties if sufficient spectra cannot be obtained to calibrate the photometric redshifts independently. 
In a forthcoming paper we will investigate into a more realistic approach including uncertainty and outliers in the redshift distributions of each individual photo-z bin, as well as the benefits of spectroscopic redshift information for a subsample of the galaxy catalogue. 

We emphasise that in the approach suggested here all considered correlations help
in 
constraining cosmological parameters. Hence, none of the contributions is regarded as a systematic signal, and the different levels of uncertainty concerning the exact form of the signals are represented by nuisance parameters. By means of the joint analysis we increase the statistical power via the cross-calibration abilities of the signals, and reduce the risk of undetected systematic effects one faces when considering signals such as cosmic shear, galaxy clustering, or lensing magnification individually.

This integrative ansatz is not limited to two-point correlations, but can be generalised to the three-point level in a straightforward way. Future surveys will provide excellent data for studying three-point correlations whose exploitation can break parameter degeneracies and improve constraints considerably \citep[e.g.][]{takada04}. Our knowledge about intrinsic alignments and galaxy bias at the three-point level is currently even more limited than in the two-point case
(however see \citealp{semboloni08}), 
so that a joint investigation of shape and number density observables, including a general parametrisation of the various bias terms, may be an appropriate way forward. However, limitations due to computational power will most likely play a dominant role in this case.

One can also think of incorporating further sets of observables into the analysis. For instance, planned surveys like Euclid will also include a spectroscopic survey of a subset of the galaxies in order to determine the underlying redshift distributions of the photo-z bins. The spectroscopic data can be used to measure baryonic acoustic oscillations (note that the acoustic peaks are not included in our galaxy number-density correlations) and galaxy peculiar velocities. The latter allow for the measurement of redshift space distortions. Similar to the case considered here, the joint analysis of galaxy number density, shape, and velocity information \citep{guzik09} will efficiently cross-calibrate nuisance terms such as the galaxy bias and tighten constraints on the cosmological model.

\begin{acknowledgements}
We would like to thank
Peter Schneider, 
Tom Kitching, 
Adam Amara, Gary Bernstein, Bhuvnesh Jain and Ofer Lahav for helpful discussions. 
BJ acknowledges support by the Deutsche Telekom Stiftung and the Bonn-Cologne Graduate School of Physics and Astronomy.
SLB thanks the Royal Society for support in the form of a University
Research Fellowship. 
\end{acknowledgements}

\bibliographystyle{aa}

\begin{thebibliography}{111}
\expandafter\ifx\csname natexlab\endcsname\relax\def\natexlab#1{#1}\fi

\bibitem[{Abdalla {et~al.}(2007)Abdalla, Amara, Capak, Cypriano, Lahav, \&
  Rhodes}]{abdalla07}
Abdalla, F.~B., Amara, A., Capak, P., {et~al.} 2007, MNRAS, 387, 969

\bibitem[{Albrecht {et~al.}(2006)Albrecht, Bernstein, Cahn, Freedman, Hewitt,
  \& et~al.}]{albrecht06}
Albrecht, A., Bernstein, G., Cahn, R., {et~al.} 2006, astro-ph/0609591

\bibitem[{Annis {et~al.}(2005)Annis, Bridle, Castander, Evrard, Fosalba,
  {et~al.}}]{annis05}
Annis, J., Bridle, S.~L., Castander, F.~J., {et~al.} 2005, astro-ph/0510195

\bibitem[{Bacon {et~al.}(2000)Bacon, R{\'e}fr{\'e}gier, \& Ellis}]{bacon00}
Bacon, D.~J., R{\'e}fr{\'e}gier, A.~R., \& Ellis, R.~S. 2000, MNRAS, 318, 625

\bibitem[{Bartelmann \& Schneider(2001)}]{bartelmann01}
Bartelmann, M. \& Schneider, P. 2001, Phys. Reports, 340, 291

\bibitem[{Benjamin {et~al.}(2007)Benjamin, Heymans, Semboloni, van Waerbeke,
  Hoekstra, more, \& more}]{benjamin07}
Benjamin, J., Heymans, C., Semboloni, E., {et~al.} 2007, MNRAS, 381, 702

\bibitem[{Bernstein(2009)}]{bernstein08}
Bernstein, G.~M. 2009, ApJ, 695, 652

\bibitem[{Bernstein \& Huterer(2010)}]{bernstein09}
Bernstein, G.~M. \& Huterer, D. 2010, MNRAS, 401, 1399

\bibitem[{Blake \& Bridle(2005)}]{blake05}
Blake, C. \& Bridle, S. 2005, MNRAS, 363, 1329

\bibitem[{Blake {et~al.}(2007)Blake, Collister, Bridle, \& Lahav}]{blake07}
Blake, C., Collister, A., Bridle, S., \& Lahav, O. 2007, MNRAS, 374, 1527

\bibitem[{Brainerd {et~al.}(2009)Brainerd, Agustsson, Madsen, \&
  Edmonds}]{brainerd09}
Brainerd, T., Agustsson, I., Madsen, C.~A., \& Edmonds, J.~A. 2009,
  astro-ph/0904.3095, submitted to ApJ

\bibitem[{Bridle \& Abdalla(2007)}]{bridle07a}
Bridle, S. \& Abdalla, F.~B. 2007, ApJ, 655, L1

\bibitem[{Bridle \& King(2007)}]{bridle07}
Bridle, S. \& King, L. 2007, NJPh, 9, 444

\bibitem[{Broadhurst {et~al.}(1995)Broadhurst, Taylor, \&
  Peacock}]{broadhurst95}
Broadhurst, T.~J., Taylor, A.~N., \& Peacock, J.~A. 1995, ApJ, 438, 49

\bibitem[{Brown {et~al.}(2002)Brown, Taylor, Hambly, \& Dye}]{brown02}
Brown, M.~L., Taylor, A.~N., Hambly, N.~C., \& Dye, S. 2002, MNRAS, 333, 501

\bibitem[{Cacciato {et~al.}(2009)Cacciato, van~den Bosch, More, Li,
  {et~al.}}]{cacciato08}
Cacciato, M., van~den Bosch, F.~C., More, S., Li, R., {et~al.} 2009, MNRAS,
  394, 929

\bibitem[{Catelan {et~al.}(2001)Catelan, Kamionkowski, \&
  Blandford}]{catelan01}
Catelan, P., Kamionkowski, M., \& Blandford, R.~D. 2001, MNRAS, 320, 7

\bibitem[{Chevallier \& Polarski(2001)}]{chevallier01}
Chevallier, M. \& Polarski, D. 2001, Int. J. Mod. Phys., D10

\bibitem[{Cresswell \& Percival(2009)}]{cresswell09}
Cresswell, J.~G. \& Percival, W.~J. 2009, MNRAS, 392, 682

\bibitem[{Crittenden {et~al.}(2001)Crittenden, Natarajan, Pen, \&
  Theuns}]{crittenden01}
Crittenden, R.~G., Natarajan, P., Pen, U., \& Theuns, T. 2001, ApJ, 559, 552

\bibitem[{Croft \& Metzler(2000)}]{croft00}
Croft, R.~A.~C. \& Metzler, C.~A. 2000, ApJ, 545, 561

\bibitem[{Das \& Spergel(2009)}]{das08}
Das, S. \& Spergel, D.~N. 2009, Phys. Rev. D, 79, 043509

\bibitem[{Dolney {et~al.}(2006)Dolney, Jain, \& Takada}]{dolney06}
Dolney, D., Jain, B., \& Takada, M. 2006, MNRAS, 366, 884

\bibitem[{Eisenstein \& Hu(1998)}]{eisenstein98}
Eisenstein, D.~J. \& Hu, W. 1998, ApJ, 496, 605

\bibitem[{Fu {et~al.}(2008)Fu, Semboloni, Hoekstra, et~al., more, \&
  more}]{fu07}
Fu, L., Semboloni, E., Hoekstra, H., {et~al.} 2008, A\&A, 479, 9

\bibitem[{Guzik {et~al.}(2010)Guzik, Jain, \& Takada}]{guzik09}
Guzik, J., Jain, B., \& Takada, M. 2010, Phys. Rev. D, 81, 023503

\bibitem[{Guzik \& Seljak(2001)}]{guzik01}
Guzik, J. \& Seljak, U. 2001, MNRAS, 321, 439

\bibitem[{Guzik \& Seljak(2002)}]{guzik02}
Guzik, J. \& Seljak, U. 2002, MNRAS, 335, 311

\bibitem[{Heavens {et~al.}(2000)Heavens, R{\'e}fr{\'e}gier, \&
  Heymans}]{heavens00}
Heavens, A., R{\'e}fr{\'e}gier, A., \& Heymans, C. 2000, MNRAS, 319

\bibitem[{Hetterscheidt {et~al.}(2007)Hetterscheidt, Simon, Schirmer,
  Hildebrandt, Schrabback, Erben, \& Schneider}]{hetterscheidt07}
Hetterscheidt, M., Simon, P., Schirmer, M., {et~al.} 2007, A\&A, 468, 859

\bibitem[{Heymans {et~al.}(2004)Heymans, Brown, Heavens, Meisenheimer, Taylor,
  \& Wolf}]{heymans04}
Heymans, C., Brown, M., Heavens, A., {et~al.} 2004, MNRAS, 347, 895

\bibitem[{Heymans \& Heavens(2003)}]{heymans03}
Heymans, C. \& Heavens, A. 2003, MNRAS, 339, 711

\bibitem[{Hirata {et~al.}(2007)Hirata, Mandelbaum, Ishak, Seljak, Nichol,
  Pimbblet, Ross, \& Wake}]{hirata07}
Hirata, C.~M., Mandelbaum, R., Ishak, M., {et~al.} 2007, MNRAS, 381, 1197

\bibitem[{Hirata \& Seljak(2004)}]{hirata04}
Hirata, C.~M. \& Seljak, U. 2004, Phys. Rev. D, 70, 063526

\bibitem[{Hoekstra {et~al.}(2006)Hoekstra, Mellier, van Waerbeke, Semboloni,
  Fu, Hudson, Parker, Tereno, \& Benabed}]{hoekstra06}
Hoekstra, H., Mellier, Y., van Waerbeke, L., {et~al.} 2006, ApJ, 647, 116

\bibitem[{Hu(1999)}]{hu99}
Hu, W. 1999, ApJ, 522, 21

\bibitem[{Hu(2002)}]{hu02b}
Hu, W. 2002, Phys. Rev. D, 65, 023003

\bibitem[{Hu \& Jain(2004)}]{hu03}
Hu, W. \& Jain, B. 2004, Phys. Rev. D, 70, 043009

\bibitem[{Huterer {et~al.}(2006)Huterer, Takada, Bernstein, \&
  Jain}]{huterer06}
Huterer, D., Takada, M., Bernstein, G., \& Jain, B. 2006, MNRAS, 366, 101

\bibitem[{Jarvis {et~al.}(2006)Jarvis, Jain, Bernstein, \& Dolney}]{jarvis06}
Jarvis, M., Jain, B., Bernstein, G., \& Dolney, D. 2006, ApJ, 644, 71

\bibitem[{Jing(2002)}]{jing02}
Jing, Y.~P. 2002, MNRAS, 335, 89

\bibitem[{Joachimi \& Schneider(2008)}]{joachimi08b}
Joachimi, B. \& Schneider, P. 2008, A\&A, 488, 829

\bibitem[{Joachimi \& Schneider(2009)}]{joachimi09}
Joachimi, B. \& Schneider, P. 2009, A\&A, 507, 105

\bibitem[{Joachimi {et~al.}(2008)Joachimi, Schneider, \& Eifler}]{joachimi08}
Joachimi, B., Schneider, P., \& Eifler, T. 2008, A\&A, 477, 43

\bibitem[{Johnston {et~al.}(2007)Johnston, Sheldon, Tasitsiomi, Frieman,
  {et~al.}}]{johnston07}
Johnston, D.~E., Sheldon, E.~S., Tasitsiomi, A., Frieman, J.~A., {et~al.} 2007,
  ApJ, 656, 27

\bibitem[{Kaiser(1992)}]{kaiser92}
Kaiser, N. 1992, ApJ, 388, 272

\bibitem[{Kaiser {et~al.}(2000)Kaiser, Wilson, \& Luppino}]{kaiser00}
Kaiser, N., Wilson, G., \& Luppino, G. 2000, astro-ph/0003338

\bibitem[{King(2005)}]{king05}
King, L.~J. 2005, A\&A, 441, 47

\bibitem[{King \& Schneider(2002)}]{king02}
King, L.~J. \& Schneider, P. 2002, A\&A, 396, 411

\bibitem[{King \& Schneider(2003)}]{king03}
King, L.~J. \& Schneider, P. 2003, A\&A, 398, 23

\bibitem[{Kitching {et~al.}(2009)Kitching, Amara, Abdalla, Joachimi, \&
  R{\'e}fr{\'e}gier}]{kitching08}
Kitching, T.~D., Amara, A., Abdalla, F.~B., Joachimi, B., \& R{\'e}fr{\'e}gier,
  A. 2009, MNRAS, 399, 2107

\bibitem[{Kitching {et~al.}(2008{\natexlab{a}})Kitching, Heavens, Verde, Serra,
  \& Melchiorri}]{kitching08b}
Kitching, T.~D., Heavens, A.~F., Verde, L., Serra, P., \& Melchiorri, A.
  2008{\natexlab{a}}, Phys. Rev. D, 77, 103008

\bibitem[{Kitching {et~al.}(2008{\natexlab{b}})Kitching, Taylor, \&
  Heavens}]{kitching08c}
Kitching, T.~D., Taylor, A.~N., \& Heavens, A.~F. 2008{\natexlab{b}}, MNRAS,
  389, 173

\bibitem[{Krause \& Hirata(2009)}]{krause09}
Krause, E. \& Hirata, C. 2009, astro-ph/0910.3786, submitted to A\&A

\bibitem[{{Lahav} \& {Suto}(2004)}]{lahav04}
{Lahav}, O. \& {Suto}, Y. 2004, Liv. Rev. Relativ., 7, 8

\bibitem[{Laureijs {et~al.}(2009)}]{laureijs09}
Laureijs, R. {et~al.} 2009, Euclid Assessment Study Report for the ESA Cosmic
  Visions, ESA/SRE(2009)2, astro-ph/0912.0914

\bibitem[{Lee \& Pen(2000)}]{lee00}
Lee, J. \& Pen, U.-L. 2000, ApJ, 532, L5

\bibitem[{Linder(2003)}]{linder03}
Linder, E.~V. 2003, Phys. Rev. Lett., 90, 091301

\bibitem[{Liu {et~al.}(2008)Liu, Capak, Mobasher, Paglione, Rich,
  {et~al.}}]{liu08}
Liu, C.~T., Capak, P., Mobasher, B., {et~al.} 2008, ApJ, 672, 198

\bibitem[{Ma {et~al.}(2006)Ma, Hu, \& Huterer}]{ma05}
Ma, Z., Hu, W., \& Huterer, D. 2006, ApJ, 636, 21

\bibitem[{Mackey {et~al.}(2002)Mackey, White, \& Kamionkowski}]{mackey02}
Mackey, J., White, M., \& Kamionkowski, M. 2002, MNRAS, 332, 788

\bibitem[{{Mandelbaum} {et~al.}(2010){Mandelbaum}, {Blake}, {Bridle},
  {Abdalla}, {et~al.}}]{mandelbaum09}
{Mandelbaum}, R., {Blake}, C., {Bridle}, S., {Abdalla}, F.~B., {et~al.} 2010,
  MNRAS, accepted

\bibitem[{Mandelbaum {et~al.}(2006)Mandelbaum, Hirata, Ishak, Seljak, \&
  Brinkmann}]{mandelbaum06}
Mandelbaum, R., Hirata, C.~M., Ishak, M., Seljak, U., \& Brinkmann, J. 2006,
  MNRAS, 367, 611

\bibitem[{Massey {et~al.}(2007)Massey, Rhodes, Leauthaud, Capak, Ellis,
  {et~al.}}]{massey07}
Massey, R., Rhodes, J., Leauthaud, A., {et~al.} 2007, ApJS, 172, 239

\bibitem[{McCracken {et~al.}(2008)McCracken, Ilbert, Mellier, Bertin,
  {et~al.}}]{mccracken07}
McCracken, H.~J., Ilbert, O., Mellier, Y., Bertin, E., {et~al.} 2008, A\&A,
  479, 321

\bibitem[{McDonald {et~al.}(2006)McDonald, Trac, \& Contaldi}]{mcdonald06}
McDonald, P., Trac, H., \& Contaldi, C. 2006, MNRAS, 366, 547

\bibitem[{Newman(2008)}]{newman08}
Newman, J.~A. 2008, ApJ, 684, 88

\bibitem[{Okumura \& Jing(2009)}]{okumura09}
Okumura, T. \& Jing, Y.~P. 2009, ApJ, 694, L83

\bibitem[{Okumura {et~al.}(2009)Okumura, Jing, \& Li}]{okumura08}
Okumura, T., Jing, Y.~P., \& Li, C. 2009, ApJ, 694, 214

\bibitem[{Padmanabhan {et~al.}(2007)Padmanabhan, Schlegel, Seljak, Makarov,
  {et~al.}}]{padmanabhan07}
Padmanabhan, N., Schlegel, D.~J., Seljak, U., Makarov, A., {et~al.} 2007,
  MNRAS, 378, 852

\bibitem[{Peacock {et~al.}(2006)Peacock, Schneider, Efstathiou, Ellis,
  Leibundgut, Lilly, \& Mellier}]{peacock06}
Peacock, J.~A., Schneider, P., Efstathiou, G., {et~al.} 2006, in ESA-ESO
  Working Group on ''Fundamental Cosmology'', ed. E.~J.~A. Peacock~et al.

\bibitem[{Pen {et~al.}(2000)Pen, Lee, \& Seljak}]{pen00}
Pen, U.-L., Lee, J., \& Seljak, U. 2000, ApJ, 543, L107

\bibitem[{Rassat {et~al.}(2008)Rassat, Amara, Amendola, Castander, Kitching, \&
  et~al.}]{rassat08}
Rassat, A., Amara, A., Amendola, L., {et~al.} 2008, astro-ph/0810.0003,
  submitted to MNRAS

\bibitem[{R{\'e}fr{\'e}gier {et~al.}(2008)R{\'e}fr{\'e}gier, Amara, Kitching,
  \& Rassat}]{refregier08}
R{\'e}fr{\'e}gier, A., Amara, A., Kitching, T., \& Rassat, A. 2008,
  astro-ph/0810.1285, submitted to A\&A

\bibitem[{R{\'e}fr{\'e}gier {et~al.}(2006)R{\'e}fr{\'e}gier, Boulade, Mellier,
  Milliard, Pain, {et~al.}}]{refregier06}
R{\'e}fr{\'e}gier, A., Boulade, O., Mellier, Y., {et~al.} 2006, SPIE, 6265,
  62651Y

\bibitem[{R{\'e}fr{\'e}gier {et~al.}(2004)R{\'e}fr{\'e}gier, Massey, Rhodes,
  Ellis, Albert, {et~al.}}]{refregier04}
R{\'e}fr{\'e}gier, A., Massey, R., Rhodes, J., {et~al.} 2004, AJ, 127, 3102

\bibitem[{Sch{\"a}fer(2009)}]{schaefer08}
Sch{\"a}fer, B.~M. 2009, IJMPD, 18, 173

\bibitem[{Schmidt {et~al.}(2009)Schmidt, Rozo, Dodelson, Hui, \&
  Sheldon}]{schmidt09}
Schmidt, F., Rozo, E., Dodelson, S., Hui, L., \& Sheldon, E. 2009, Phys. Rev.
  Lett., 103, 051301

\bibitem[{Schneider {et~al.}(2006)Schneider, Knox, Zhan, \&
  Connolly}]{mschneider06}
Schneider, M., Knox, L., Zhan, H., \& Connolly, A. 2006, ApJ, 651, 14

\bibitem[{Schneider \& Bridle(2010)}]{schneiderm09}
Schneider, M.~D. \& Bridle, S. 2010, MNRAS, 402, 2127

\bibitem[{Schneider(2006)}]{schneider06}
Schneider, P. 2006, in Saas-Fee Advanced Course 33: Gravitational Lensing:
  Strong, Weak and Micro, ed. G.~{Meylan}, P.~{Jetzer}, P.~{North},
  P.~{Schneider}, C.~S. {Kochanek}, \& J.~{Wambsganss}, 269

\bibitem[{Schneider \& Rix(1997)}]{schneider97}
Schneider, P. \& Rix, H.-W. 1997, ApJ, 474, 25

\bibitem[{Schrabback {et~al.}(2007)Schrabback, Erben, Simon, Miralles,
  Schneider, \& et~al.}]{schrabback07}
Schrabback, T., Erben, T., Simon, P., {et~al.} 2007, A\&A, 468, 823

\bibitem[{Schrabback {et~al.}(2010)Schrabback, Hartlap, Joachimi, Kilbinger,
  Simon, {et~al.}}]{schrabback09}
Schrabback, T., Hartlap, J., Joachimi, B., {et~al.} 2010, A\&A, 516, 63

\bibitem[{Seitz \& Schneider(1997)}]{seitz97}
Seitz, C. \& Schneider, P. 1997, A\&A, 318, 687

\bibitem[{Seljak(2002)}]{seljak02}
Seljak, U. 2002, MNRAS, 337, 769

\bibitem[{Seljak {et~al.}(2005)Seljak, Makarov, Mandelbaum, Hirata,
  {et~al.}}]{seljak05}
Seljak, U., Makarov, A., Mandelbaum, R., Hirata, C.~M., {et~al.} 2005, Phys.
  Rev. D, 71, 043511

\bibitem[{Semboloni {et~al.}(2008)Semboloni, Heymans, van Waerbeke, \&
  Schneider}]{semboloni08}
Semboloni, E., Heymans, C., van Waerbeke, L., \& Schneider, P. 2008, MNRAS,
  388, 991

\bibitem[{Semboloni {et~al.}(2006)Semboloni, Mellier, van Waerbeke, Hoekstra,
  Tereno, Benabed, Gwyn, Fu, Hudson, Maoli, \& Parker}]{semboloni06}
Semboloni, E., Mellier, Y., van Waerbeke, L., {et~al.} 2006, A\&A, 452, 51

\bibitem[{Simon {et~al.}(2009)Simon, Hetterscheidt, Wolf, Meisenheimer,
  Hildebrandt, {et~al.}}]{simon08}
Simon, P., Hetterscheidt, M., Wolf, C., {et~al.} 2009, MNRAS, 398, 807

\bibitem[{Simon {et~al.}(2004)Simon, King, \& Schneider}]{simon04}
Simon, P., King, L.~J., \& Schneider, P. 2004, A\&A, 417, 873

\bibitem[{Smail {et~al.}(1994)Smail, Ellis, \& Fitchett}]{smail94}
Smail, I., Ellis, R.~S., \& Fitchett, M. 1994, MNRAS, 270, 245

\bibitem[{Smith {et~al.}(2003)Smith, Peacock, Jenkins, White, Frenk, Pearce,
  Thomas, Efstathiou, \& Couchman}]{smith03}
Smith, R.~E., Peacock, J.~A., Jenkins, A., {et~al.} 2003, MNRAS, 341, 1311

\bibitem[{Swanson {et~al.}(2008)Swanson, Tegmark, Blanton, \&
  Zehavi}]{swanson08}
Swanson, M. E.~C., Tegmark, M., Blanton, M., \& Zehavi, I. 2008, MNRAS, 385,
  1635

\bibitem[{Takada \& Jain(2004)}]{takada04}
Takada, M. \& Jain, B. 2004, MNRAS, 348, 897

\bibitem[{Takada \& White(2004)}]{takada04b}
Takada, M. \& White, M. 2004, ApJ, 601, 1

\bibitem[{Tegmark {et~al.}(1997)Tegmark, Taylor, \& Heavens}]{tegmark97}
Tegmark, M., Taylor, A.~N., \& Heavens, A.~F. 1997, ApJ, 480, 22

\bibitem[{Tereno {et~al.}(2005)Tereno, Dor{\'e}, van Waerbeke, \&
  Mellier}]{tereno05}
Tereno, I., Dor{\'e}, O., van Waerbeke, L., \& Mellier, Y. 2005, A\&A, 429, 383

\bibitem[{Thomas {et~al.}(2009)Thomas, Abdalla, \& Weller}]{thomas08}
Thomas, S.~A., Abdalla, F.~B., \& Weller, J. 2009, MNRAS, 395, 197

\bibitem[{van~den Bosch {et~al.}(2002)van~den Bosch, Abel, Croft, Hernquist, \&
  White}]{bosch02}
van~den Bosch, F.~C., Abel, T., Croft, R.~A.~C., Hernquist, L., \& White,
  S.~D.~M. 2002, ApJ, 576, 21

\bibitem[{van Waerbeke(2010)}]{vwaer09}
van Waerbeke, L. 2010, MNRAS, 401, 2093

\bibitem[{van Waerbeke {et~al.}(2000)van Waerbeke, Mellier, Erben, Cuillandre,
  Bernardeau, Maoli, Bertin, McCracken, Fevre, Fort, Dantel-Fort, Jain, \&
  Schneider}]{vwaer00}
van Waerbeke, L., Mellier, Y., Erben, T., {et~al.} 2000, A\&A, 358, 30

\bibitem[{Wang {et~al.}(2007)Wang, Yang, Mo, \& van~den Bosch}]{wang07}
Wang, Y., Yang, X.-H., Mo, H.~J., \& van~den Bosch, F.~C. 2007, ApJ, 664, 608

\bibitem[{Wittman {et~al.}(2000)Wittman, Tyson, Kirkman, Antonio, \&
  Bernstein}]{wittman00}
Wittman, D.~M., Tyson, J.~A., Kirkman, D., Antonio, I.~D., \& Bernstein, G.
  2000, Nature, 405, 143

\bibitem[{Wolf {et~al.}(2003)Wolf, Meisenheimer, Rix, Borch, Dye, \&
  Kleinheinrich}]{wolf03}
Wolf, C., Meisenheimer, K., Rix, H.-W., {et~al.} 2003, A\&A, 401, 73

\bibitem[{Yoo {et~al.}(2006)Yoo, Tinker, Weinberg, Zheng, Katz,
  {et~al.}}]{yoo06}
Yoo, J., Tinker, J.~L., Weinberg, D.~H., {et~al.} 2006, ApJ, 652, 26

\bibitem[{Zhan(2006)}]{zhan06}
Zhan, H. 2006, JCAP, 0608, 008

\bibitem[{Zhang(2010)}]{zhang08}
Zhang, P. 2010, ApJ, accepted

\bibitem[{Zhang \& Pen(2005)}]{zhang05}
Zhang, P. \& Pen, U.-L. 2005, Phys. Rev. Lett., 95, 241302

\bibitem[{Zhang \& Pen(2006)}]{zhang06}
Zhang, P. \& Pen, U.-L. 2006, MNRAS, 367, 169

\bibitem[{Zhang {et~al.}(2010)Zhang, Pen, \& Bernstein}]{zhang09}
Zhang, P., Pen, U.-L., \& Bernstein, G. 2010, MNRAS, 405, 359

\end{thebibliography}

\begin{appendix}

\section{Projected lensing magnification signal}
\label{app:magnification}

This appendix details the derivation of the enhancement or depletion of projected galaxy counts $n^{(i)}_{\rm m}(\vek{\theta})$ due to lensing magnification. The number density of galaxies $n$, counted above a flux threshold $S$ at angular position $\vek{\theta}$ and comoving distance $\chi$ is altered by gravitational lensing according to
\eq{
\label{eq:magbias}
n(>S,f_{\rm K}(\chi) \vek{\theta}, \chi) = \frac{1}{\mu(\vek{\theta}, \chi)}\; n_0 \br{> \frac{S}{\mu(\vek{\theta}, \chi)},f_{\rm K}(\chi) \vek{\theta}, \chi }\;,
}
where $n_0$ is the original galaxy number density, and where $\mu$ denotes the magnification \citep{bartelmann01}. One assumes that the galaxy luminosity function close to the flux limit of the survey can locally be written as a power law, $n(>S,f_{\rm K}(\chi) \vek{\theta}, \chi) \propto S^{- \alpha(\chi)}$. The slope $\alpha$ depends on the line-of-sight distance, or equivalently, redshift, but should not depend on angular dimensions due to isotropy. However, it is a function of the magnitude limit in the observed filter, in this work denoted by $r_{\rm lim}$. This dependence is dealt with in Sect.$\,$\ref{sec:alphalum}, but for ease of notation we drop $r_{\rm lim}$ as an argument of $\alpha$ for the remainder of this section. Plugging the power-law form of the luminosity function into (\ref{eq:magbias}) yields
\eq{
\label{eq:countratio}
\frac{n(>S,f_{\rm K}(\chi) \vek{\theta}, \chi)}{n_0(>S,f_{\rm K}(\chi) \vek{\theta}, \chi)} = \mu(\vek{\theta}, \chi)^{\alpha(\chi)-1}\;.
}

Again following \citet{bartelmann01}, one can approximate the magnification in the weak lensing regime as \mbox{$\mu \approx 1 + 2\kappa_{\rm G}$}. Since $\kappa_{\rm G} \ll 1$, we can in addition do a Taylor approximation to arrive at
\eqa{
\label{eq:countratio2}
\frac{n(>S,f_{\rm K}(\chi) \vek{\theta}, \chi)}{n_0(>S,f_{\rm K}(\chi) \vek{\theta}, \chi)} &\approx& \br{ 1 + 2\; \kappa_{\rm G}(\vek{\theta}, \chi) }^{\alpha(\chi)-1}\\ \nn
&\approx& 1 + 2\; (\alpha(\chi)-1)\; \kappa_{\rm G}(\vek{\theta}, \chi)\;.
}
Defining the excess galaxy density contrast due to magnification effects as
\eqa{
\label{eq:magbiascontrast}
\delta^{\rm m}_{\rm g}(f_{\rm K}(\chi) \vek{\theta}, \chi) &\equiv& \frac{n(>S,f_{\rm K}(\chi) \vek{\theta}, \chi)}{n_0(>S,f_{\rm K}(\chi) \vek{\theta}, \chi)} -1\\ \nn
&=& 2\; (\alpha(\chi)-1)\; \kappa_{\rm G}(\vek{\theta}, \chi)\;,
}
one obtains for the corresponding projected density contrast
\eqa{
\label{eq:magbiasprojected}
n^{(i)}_{\rm m}(\vek{\theta}) &=& \int^{\chi_{\rm hor}}_0 \dd \chi\; p^{(i)}(\chi)\;  \delta^{\rm m}_{\rm g} \br{ f_{\rm K}(\chi) \vek{\theta}, \chi }\\ \nn
&=& \int^{\chi_{\rm hor}}_0 \dd \chi\; p^{(i)}(\chi)\; 2\; (\alpha(\chi)-1)\; \kappa_{\rm G}(\vek{\theta},\chi)\;.
}
In exact analogy to the standard derivation of (\ref{eq:projectionlensing}) one can now insert the relation between the convergence and the three-dimensional matter density contrast,
\eq{
\label{eq:kappawithchi}
\kappa_{\rm G}(\vek{\theta},\chi) = \frac{3 H_0^2 \Omega_{\rm m}}{2 c^2}\!\! \int_0^\chi \!\! \dd \chi'\; \frac{f_{\rm k}(\chi') ~f_{\rm k}(\chi-\chi')}{f_{\rm k}(\chi)} ~\frac{\delta\br{f_{\rm k}(\chi') \vek{\theta},\chi'}}{a(\chi')}\;,
}
which, after swapping the order of integration and the names of the integration variables, yields
\eq{
\label{eq:magbiasprojected2}
n^{(i)}_{\rm m}(\vek{\theta}) = \int^{\chi_{\rm hor}}_0 \dd \chi\; \bar{q}^{(i)}(\chi)\; \delta \br{ f_{\rm K}(\chi) \vek{\theta}, \chi }\;.
}
Here, we have defined the weight
\eqa{
\label{eq:weightmagbias}
\bar{q}^{(i)}(\chi) &=& \frac{3 H_0^2 \Omega_{\rm m}}{2\, c^2} \frac{f_{\rm K}(\chi)}{a(\chi)} \int_{\chi}^{\chi_{\rm hor}} \dd \chi'\; p^{(i)}(\chi')\;\\ \nn
&& \hspace*{2cm} \times\; \frac{f_{\rm K}(\chi' - \chi)}{f_{\rm K}(\chi')}\; 2\,(\alpha(\chi')-1)\;.
}

Given that the slope of the luminosity function should be a smooth function of comoving distance, $\alpha(\chi)$ varies only weakly over the range of the integration in (\ref{eq:weightmagbias}), being determined by the distribution $p^{(i)}(\chi)$, which has relatively compact support. Hence, the mean value theorem constitutes a good approximation, so that we can write
\eq{
\label{eq:weightmagbiassimplified}
\bar{q}^{(i)}(\chi) \approx 2\,(\alpha^{(i)}-1)\; q^{(i)}(\chi)\;,
}
where we define $\alpha^{(i)}$ to be the slope of the luminosity function, evaluated at the median redshift of the photometric bin $i$. Inserting (\ref{eq:weightmagbiassimplified}) into (\ref{eq:magbiasprojected2}) results in (\ref{eq:projectionmag}), which we employ throughout this work.

\section{Limber equations of cosmological signals}
\label{app:limber}

For reference, we have collected here the Fourier-space Limber equations of all projected power spectra that contribute to the cosmological signals considered in this work:
\eqa{
\label{eq:limberequations}
C^{(ij)}_{\rm GG}(\ell) &=& \int^{\chi_{\rm hor}}_0 \dd \chi\; \frac{q^{(i)}(\chi)\; q^{(j)}(\chi)}{f^2_{\rm K}(\chi)}\; P_{\delta \delta} \br{\frac{\ell}{f_{\rm K}(\chi)},\chi}\;\\
C^{(ij)}_{\rm IG}(\ell) &=& \int^{\chi_{\rm hor}}_0 \dd \chi\; \frac{p^{(i)}(\chi)\; q^{(j)}(\chi)}{f^2_{\rm K}(\chi)}\; P_{\delta {\rm I}} \br{\frac{\ell}{f_{\rm K}(\chi)},\chi}\;\\
C^{(ij)}_{\rm II}(\ell) &=& \int^{\chi_{\rm hor}}_0 \dd \chi\; \frac{p^{(i)}(\chi)\; p^{(j)}(\chi)}{f^2_{\rm K}(\chi)}\; P_{\rm II} \br{\frac{\ell}{f_{\rm K}(\chi)},\chi}\;\\
C^{(ij)}_{\rm gg}(\ell) &=& \int^{\chi_{\rm hor}}_0 \dd \chi\; \frac{p^{(i)}(\chi)\; p^{(j)}(\chi)}{f^2_{\rm K}(\chi)}\; P_{\rm gg} \br{\frac{\ell}{f_{\rm K}(\chi)},\chi}\;\\
C^{(ij)}_{\rm gm}(\ell) &=& \int^{\chi_{\rm hor}}_0 \dd \chi\; \frac{p^{(i)}(\chi)\; \bar{q}^{(j)}(\chi)}{f^2_{\rm K}(\chi)}\; P_{{\rm g} \delta} \br{\frac{\ell}{f_{\rm K}(\chi)},\chi}\;\\
C^{(ij)}_{\rm mm}(\ell) &=& \int^{\chi_{\rm hor}}_0 \dd \chi\; \frac{\bar{q}^{(i)}(\chi)\; \bar{q}^{(j)}(\chi)}{f^2_{\rm K}(\chi)}\; P_{\delta \delta} \br{\frac{\ell}{f_{\rm K}(\chi)},\chi}\;\\
C^{(ij)}_{\rm gG}(\ell) &=& \int^{\chi_{\rm hor}}_0 \dd \chi\; \frac{p^{(i)}(\chi)\; q^{(j)}(\chi)}{f^2_{\rm K}(\chi)}\; P_{{\rm g} \delta} \br{\frac{\ell}{f_{\rm K}(\chi)},\chi}\;\\
C^{(ij)}_{\rm gI}(\ell) &=& \int^{\chi_{\rm hor}}_0 \dd \chi\; \frac{p^{(i)}(\chi)\; p^{(j)}(\chi)}{f^2_{\rm K}(\chi)}\; P_{\rm gI} \br{\frac{\ell}{f_{\rm K}(\chi)},\chi}\;\\
C^{(ij)}_{\rm mG}(\ell) &=& \int^{\chi_{\rm hor}}_0 \dd \chi\; \frac{\bar{q}^{(i)}(\chi)\; q^{(j)}(\chi)}{f^2_{\rm K}(\chi)}\; P_{\delta \delta} \br{\frac{\ell}{f_{\rm K}(\chi)},\chi}\;\\
C^{(ij)}_{\rm mI}(\ell) &=& \int^{\chi_{\rm hor}}_0 \dd \chi\; \frac{\bar{q}^{(i)}(\chi)\; p^{(j)}(\chi)}{f^2_{\rm K}(\chi)}\; P_{\delta {\rm I}} \br{\frac{\ell}{f_{\rm K}(\chi)},\chi}\;.
}
The terminology of both projected and three-dimensional power spectra is summarised in Table \ref{tab:correlations}. The weights that enter the foregoing equations are the probability distribution of galaxies with comoving distance $p^{(i)}(\chi)$, and the ones defined in (\ref{eq:weightlensing}), and (\ref{eq:weightmagbias}).

\section{The slope of the galaxy luminosity function}
\label{app:gallum}

\begin{figure}[t]
\centering
\includegraphics[scale=.6,angle=270]{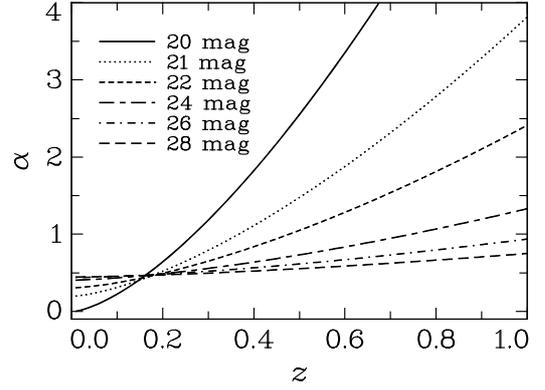}
\includegraphics[scale=.6,angle=270]{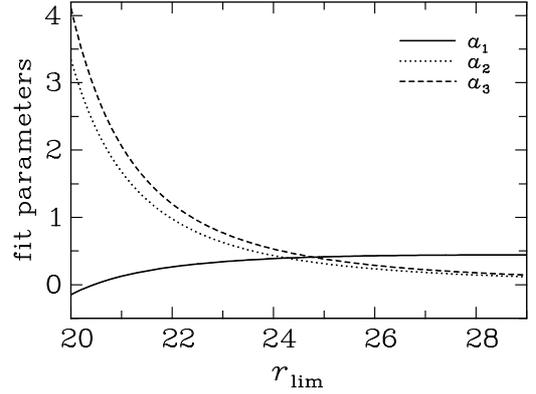}
\caption{\textit{Top panel}: Slope of the galaxy luminosity function $\alpha$ as a function of redshift, shown for different magnitude limits $r_{\rm lim}$ as indicated in the legend. \textit{Bottom panel}: Fit parameters $a_1$ (solid), $a_2$ (dotted), and $a_3$ (dashed) as a function of $r_{\rm lim}$. These parameters are obtained from the polynomial fit in (\ref{eq:alphafit1}).}
\label{fig:plot_alphas}
\end{figure}

To compute the lensing magnification signal, we need to model the power-law slope of the cumulative galaxy luminosity function at the magnitude limit of the galaxy number density samples under consideration. We base our modelling on \citet{blake05} who have determined galaxy redshift distributions for a given magnitude limit, using COMBO-17 luminosity functions for the SDSS $r$ filter \citep{wolf03}. 

They fitted these distributions with functions of the form (\ref{eq:redshiftdistribution}) with $\beta=1.5$, using two free parameters, the redshift scaling $\bar{z}$ and the normalisation given by the galaxy surface density $\Sigma_0$. We set $\Sigma_0$ and $\bar{z}$ as a function of survey depth making use of Table 1 of \citet{blake05} and fit a power law to each quantity as a function of the limiting magnitude $r_{\rm lim}$,
\eqa{
\label{eq:alphafit0}
\Sigma_0 &=& \Sigma_{0,\,c}\; {\br{\frac{r_{\rm lim}}{24}}}^{\eta_{\Sigma}}\;\\ \nn
\bar{z} &=& \bar{z}_c\, +\, \bar{z}_m\; {\br{r_{\rm lim}-24}}\;
}
where we find good fits using $\Sigma_{0,\,c}=9.83$, $\eta_{\Sigma}=19$, $\bar{z}_c=0.39$ and $\bar{z}_m=0.055$. This allows us to extrapolate beyond the range of their Table, which stops at $r_{\rm lim}=24$. 

Note that our definition of $\alpha$ is not the same as the exponent of the Schechter function (sometimes also denoted $\alpha$, see e.g. \citealp{wolf03}, eq.$\,$4). Our $\alpha$ is the negative of the slope of the cumulative luminosity function. Therefore for comparison to the Schechter function, in the faint galaxy (power-law) regime, one must take the negative of our alpha and subtract unity. Note that our use of the cumulative luminosity function is consistent with the literature on lensing magnification \citep[e.g.][]{schmidt09}. Our typical $\alpha$ values of around 0.5 are also therefore consistent with the luminosity function literature (e.g. \citealp{liu08} find Schechter function exponent values between $-1$ and $-2$ depending on the spectral type from COSMOS).

We are interested in the slope of the luminosity function $\alpha(z,r_{\rm lim})$ at the cosmic shear survey magnitude limit. This slope is a function of redshift and magnitude limit. From (\ref{eq:redshiftdistribution}) and (\ref{eq:alphafit0}) we have the number of galaxies as a function of redshift and magnitude limit. We convert each magnitude limit into a flux limit $S$ and set the number of galaxies above the flux limit equal to $S^{-\alpha}$ (see also Appendix \ref{app:magnification}). The resulting curves for $\alpha(z,r_{\rm lim})$ are shown in Fig.$\,$\ref{fig:plot_alphas}, upper panel.

\begin{table}[t]
\caption{Fit parameters for the slope of the luminosity function as a function of limiting magnitude $r_{\rm lim}$ and redshift, see (\ref{eq:alphafit1}) and (\ref{eq:alphafit2}).}
\centering
\begin{tabular}[t]{cccc}
\hline\hline
j & $b_{1\,j} $ & $b_{2\,j}$ & $b_{3\,j}$\\ 
\hline
1 & 0.44827 & 0 & 0\\
2 & -1 & +1 & +1\\
3 & 0.05617 & 0.19658 & 0.18107\\
4 & 0.07704 & 3.31359 & 3.05213\\
5 & -11.3768 & -2.5028 & -2.5027\\
\hline
\end{tabular}
\label{tab:lumfit}
\end{table}

For convenience, and to extrapolate the slope $\alpha(z,r_{\rm lim})$ to values $z>1$, we now provide a fitting formula . First we expand the slope using a polynomial in redshift, with coefficients that depend on the limiting magnitude. Then we find an approximate equation for these coefficients as a function of limiting magnitude. This results in equations for the slope as a function of redshift and magnitude limit in terms of 15 numbers given in Table \ref{tab:lumfit}. 

We fit the slope of the luminosity function as a function of redshift with a second-order polynomial
\eq{
\label{eq:alphafit1}
\alpha(z,r_{\rm lim}) = a_1 (r_{\rm lim}) + a_2(r_{\rm lim})\, z + a_3(r_{\rm lim})\, z^2\;. 
}
The polynomial coefficients $a_i$ are functions of the limiting magnitude, and are shown in Fig.$\,$\ref{fig:plot_alphas}, lower panel. We find that these coefficients are in turn well fit by a function of the form
\eq{
\label{eq:alphafit2}
a_i(r_{\rm lim}) = b_{i\,1} + b_{i\,2}\, \br{b_{i\,3}\, r_{\rm lim} - b_{i\,4}}^{b_{i\,5}} \;, 
}
with parameters $b_{i\,j}$ given in Table~\ref{tab:lumfit}. We chose not to use $b_{i\,2}$ as a free parameter for the fit, but set it as  $b_{i\,2} = \pm 1$, to determine the sign of the term in parentheses. By means of (\ref{eq:alphafit1}) and (\ref{eq:alphafit2}) we have condensed the dependence of $\alpha$ on redshift and $r_{\rm lim}$ into the 15 parameters summarised in Table \ref{tab:lumfit}.

\begin{figure}[t]
\centering
\includegraphics[scale=.6,angle=270]{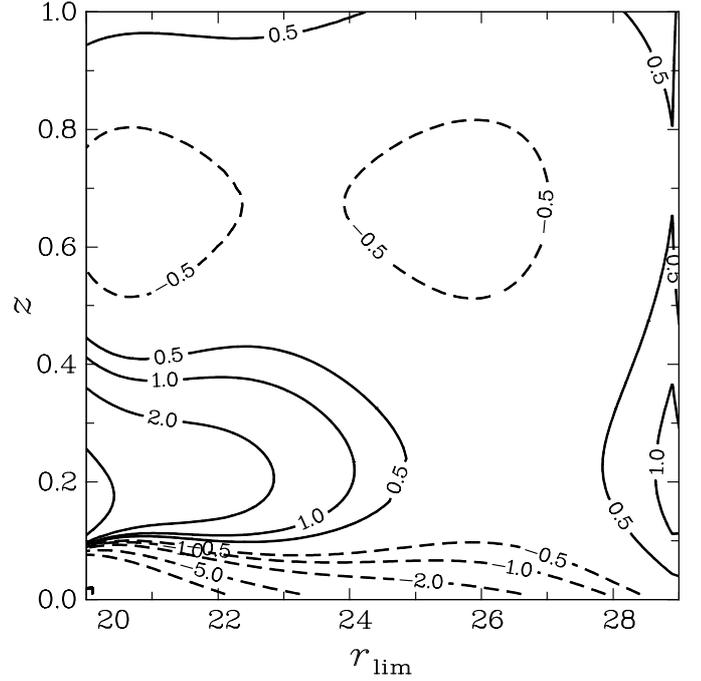}
\caption{Percentage deviation of the simplified fit as defined in (\ref{eq:alphafit1}) and (\ref{eq:alphafit2}) from the original fits described by (\ref{eq:alphafit0}) and shown in the upper panel of Fig.$\,$\ref{fig:plot_alphas}. The contour levels correspond to the percentages given on the curves. Negative deviations are indicated by dashed contours. Note that across most of the parameter plane the modulus of the deviation is less than $1\,\%$.}
\label{fig:plot_alphacontours}
\end{figure}

In Fig.$\,$\ref{fig:plot_alphacontours} we plot the relative accuracy of this set of fit formulae with respect to $\alpha(z,r_{\rm lim})$ as given in Fig.$\,$\ref{fig:plot_alphas}, lower panel. Over the dominant part of the considered parameter space the fit formulae provide an excellent approximation, which deviates less than $1\,\%$ from the original fits (\ref{eq:alphafit0}). Thus one can expect that within the framework of this approach (\ref{eq:alphafit1}) and (\ref{eq:alphafit2}) extrapolate $\alpha(z,r_{\rm lim})$ reasonably well to $z>1$. Significantly larger deviations can only be found for the brightest limiting magnitudes at redshifts $z \lesssim 0.1$, a region of the parameter plane which is irrelevant for a competitive cosmological survey.

We have used and extrapolated the COMBO17 luminosity functions despite the incompleteness beyond $R=23$ and relative unreliability in the redshift range $1.2<z<2$ due to the lack of spectral features in the observing bands used. This could be improved by using deeper data which have infra-red observations, for example COSMOS ground and space data and CFHTLS-Wide. We emphasise that this is simply used for our choice of fiducial model, and within the modelling it is assumed that the luminosity function slopes are an unknown function of redshift and are marginalised over. We do not expect the choice of fiducial model to have a large effect on our analysis. Note that the number of galaxies as a function of redshift is inserted into the analysis using the assumptions made by the project teams and is not linked to our luminosity function calculations.

It would be more elegant to simultaneously derive the redshift distribution and luminosity function slope from luminosity functions derived from data as a function of redshift. However this would make it difficult to compare our results with those in the literature from the imaging survey project teams, who have already chosen a galaxy redshift distribution. We recommend that the project teams also make available luminosity function slopes to be used in analyses such as these.

\section{Modelling the effect of dark energy on non-linear growth}
\label{sec:defitting}

In this appendix we summarise the approach made in the publicly available {\tt icosmo} code\footnote{{\tt http://www.icosmo.org}} \citep{refregier08} to model the non-linear evolution of structure in presence of a dark energy equation of state $w(z) \neq -1$, which we have adopted for this work. It is based upon a modification of the halofit routine of \citet{smith03} in which the interpolation between open and flat cosmological models is determined by the parameter $f = \Omega_\Lambda / (1 - \Omega_{\rm m})$, where $f=0$ corresponds to open universes without dark energy (we assume $\Omega_{\rm m}<1$ here), and $f=1$ to flat $\Lambda$CDM models. 

Now the fact is exploited that certain variable dark energy models mimic the expansion history of open CDM universes. Using $w_0=-1/3$ and $w_a=0$ in (\ref{eq:darkenergy}) for a flat $\Lambda$CDM model, it is readily seen that the same Hubble parameter is obtained as for an open CDM universe with identical $\Omega_{\rm m}$ but without dark energy, where (\ref{eq:darkenergy}) plays the role of the curvature term. Motivated by this coincidence, the interpolation parameter $f$ in the halofit routine is replaced by
\eq{
\label{eq:newhalofit}
f' \equiv - \frac{1}{2} \br{3 \bb{w(z)\;f - \frac{1}{3}\; (1-f)} + 1}\;,
}
where $w(z)=w_0 + w_a z/(1+z)$, and $f = \Omega_\Lambda / (1 - \Omega_{\rm m})$ as before. If a model does not feature dark energy, $f'=f=0$. For a flat $\Lambda$CDM model the interpolation now takes place between $w(z)=-1/3$, mimicking the case of an open CDM in the original halofit ($f'=0$), and the cosmological constant $w(z) \equiv -1$ ($f'=1$).

The performance of this simplistic ansatz has been tested in Fig.$\,$10 of \citet{schrabback09}, finding fair agreement with the fit formula of \citet{mcdonald06}. Note that the fits to the simulations by \citet{mcdonald06}, which include a dependence on $w_0$, are not suitable for direct use in this article due to the limited range in cosmological parameters, most notably $\sigma_8$.

\end{appendix}

\end{document}